# Real-time Object Detection and Associated Hardware Accelerators Targeting Autonomous Vehicles: A Review

Safa Sali, Anis Meribout, Ashiyana Majeed, Mahmoud Meribout, Juan Pablo, Varun Tiwari, and Asma Baobaid

*Computer Engineering & Information Engineering Department, College of Computer and Mathematical Sciences, Khalifa University of Science & Technology, Abu Dhabi, United Arab Emirates*

## Abstract

The efficiency of object detectors depends on factors like detection accuracy, processing time, and computational resources. Processing time is crucial for real-time applications, particularly for autonomous vehicles (AVs), where instantaneous responses are vital for safety. This review paper provides a concise yet comprehensive survey of real-time object detection (OD) algorithms for autonomous cars delving into their hardware accelerators (HAs). Non-neural network-based algorithms, which use statistical image processing, have been entirely substituted by AI algorithms, such as different models of convolutional neural networks (CNNs). Their intrinsically parallel features led them to be deployable into edge-based HAs of various types, where GPUs and, to a lesser extent, ASIC (application-specific integrated circuit) remain the most widely used. Throughputs of hundreds of frames/s (fps) could be reached; however, handling object detection for all the cameras available in a typical AV requires further hardware and algorithmic improvements. The intensive competition between AV providers prevented the disclosure of associated algorithms, firmware, and sometimes even hardware platform details. This constitutes a hurdle for researchers as commercial systems offer valuable input to researchers as they go through intensive and time-consuming training and tests on different roads. Consequently, thousands of research papers on AVs may not be endorsed in the end products as they were mainly developed based on existing datasets in limited situations. This paper highlights the state-of-the-art algorithms used for OD and intends to narrow the gap with technologies used in commercially available AV. To our knowledge, this aspect was not addressed in previous survey papers. Therefore, this paper can be a tangible reference for researchers designing future generations of vehicles, which are expected to be fully autonomous for comfort and safety.

*Keywords*: Autonomous vehicle, edge device, feature extraction, object detection, real-time.

## 1. Introduction

After several years of reluctance, all major car manufacturers consider AVs (i.e., cars, robotaxis, and trucks) their key product. In addition to offering drivers comfort, they mainly seek to enhance road safety, where not less than 1.3 million deaths are caused by car accidents [1]. For instance, Tesla's success in leading the electric vehicles market is losing momentum with aggressive competition from Chinese electric car manufacturers. However, the company relies on AV business to regain leadership as the Chinese competitors are exposed to US export restrictions of powerful edge processors required in AVs. This was motivated by the emergence of specialized HAs that perform real-time OD at hundreds of fps. Indeed, real-time OD remains the key enabling requirement in AVs. Currently, Tesla leads the AV market by offering fully autonomous AVs that can operate in some cities within the USA [1]. The company's most recent autopilot system is based on a customized HA, namely HW4 [2], to perform mainly two-dimensional (2D) OD on frames captured from up to 8 cameras. Recently, Toyota announced its partnership with Nvidia to build its next-generation AV system based on Nvidia's Drive AGX Orin GPU, which is more powerful than the Tesla HW4 platform [3]. Indeed, the AVs' share market is expected to reach Trillions of dollars by 2030, which will lead to AI advances as a real business value. AVs are generally equipped with multiple cameras, in addition to some sensors such as GPS (global positioning system) and ultrasonic and radar sensors to capture the current vehicle position, nearby obstacles, and their speed, respectively. Some AV manufacturers, such as Waymo, use an additional LIDAR system to perform 3D OD. Depending on the accuracy and throughput of their OD, AVs are categorized from level 1 to level 5, with level 5 featuring full autonomy that does not require human intervention. However, despite some claims, level 5 autonomous cars are still not authorized to operate, mainly due to the inability of the current OD module to cope with all possible scenarios, mostly lousy weather conditions, low light intensity, and unforeseen objects that may appear on the roads. This has led AV car manufacturers to extend the dataset for training.

Some survey papers on object detection targeting AS were recently published. For instance, in [4], authors provided various 3D object detection targeting autonomous vehicles using cameras and LIDAR systems simultaneously. They recognized the importance of using LIDAR and cameras to accurately build the environment surrounding autonomous vehicles. However, leading AV manufacturers have omitted the LIDAR system from their cars because of their high cost and the ability of camera-based systems to provide depth information from monocular images. In addition, the paper does not mention the underlying HAs used and does not tend to narrow the gap between academia and commercialized AV systems, which is the case of this review. In summary, the main contributions of this paper are:

1. Survey the most recent AI algorithms for OD targeting AVs and their associated HAs. It also highlights the potential of generative AI to enhance both the accuracy and throughput of existing OD algorithms
2. Highlight the algorithmic and hardware features exhibited in some commercial AVs to narrow the gap for academic research where real-life constraints were not considered rigorously.
3. List the challenges AV researchers face from academia and industry and suggest tangible solutions and future work opportunities.

## 2. Background

### 2.1 Aim of Object Detection for AVs

Fig. 1 illustrates the aim of OD for AVs. It simultaneously captures the frames from multiple cameras, which feature various fields of view (FOVs) and are located at different locations within a car to ensure the full coverage of the areas surrounding the vehicle. For instance, Tesla's autopilot system comprises eight cameras of two possible resolutions (2896 x 1876) and (1448 x 938) and no Lidar, bringing down its cost to around 15,000 US$ [1][2]. The three front cameras, which are the most critical, have different FOVs, as the current OD algorithm may fail when the object in the image space comprises only a few pixels. The front cameras have a range of view of 60 m, 150 m, and 250 m, respectively.

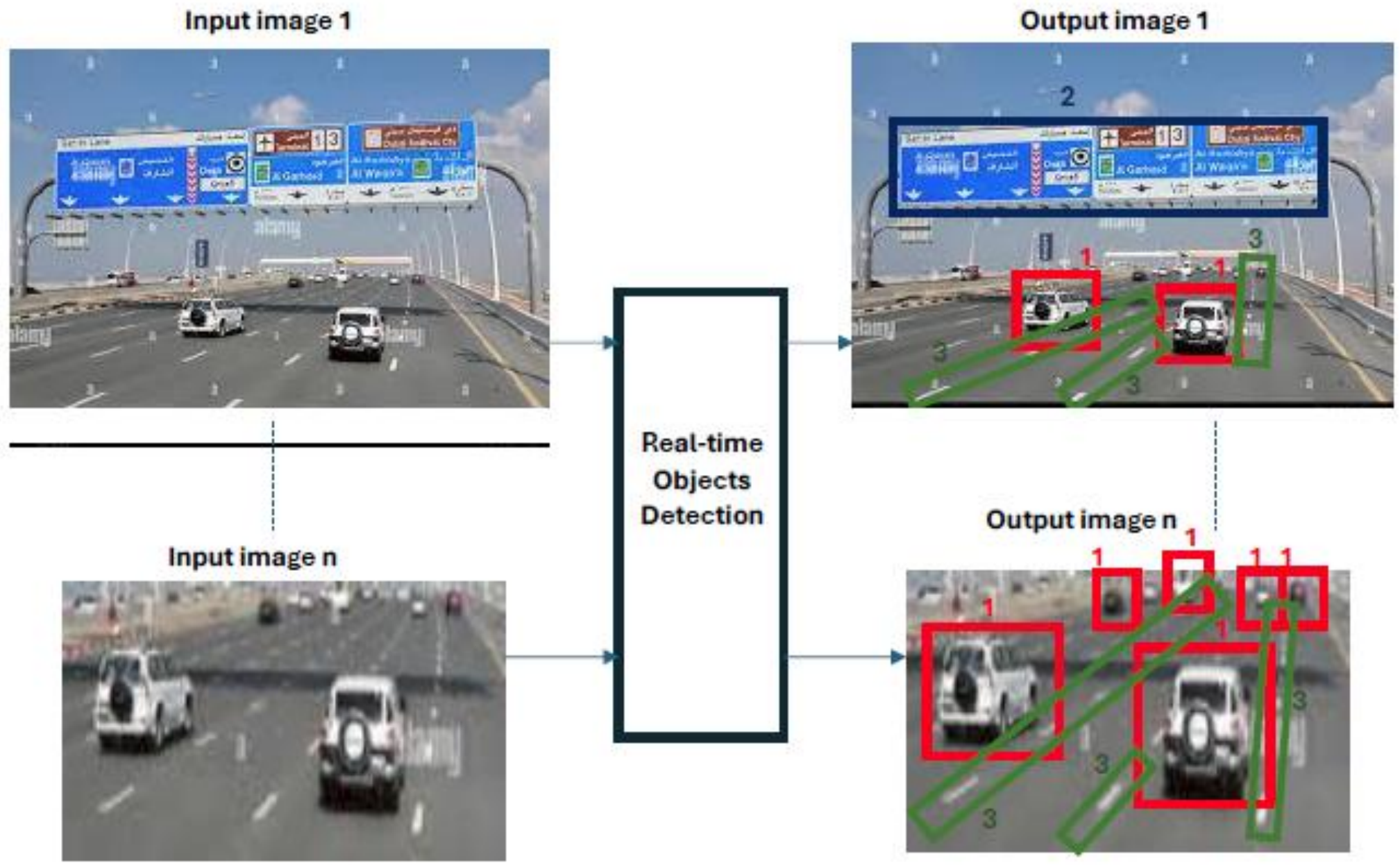

**Fig. 1** Principle of object detection task in AVs.

In addition, in each camera, several objects are ought to be detected (e.g., vehicles, pedestrians, traffic signs, traffic lights, and bicycles). This requires a high computation capability of the HAs and a proper hardware-software codesign, which researchers did not address. Indeed, cases of a single object captured from a single camera were mainly considered in the literature. One patented approach that Tesla suggests for reducing the computation requirement is to detect the region of interest (ROI) within an image and perform object detection solely on those regions [5]. The ROI may be cropped to yield higher image resolution and increase the detection accuracy. Thus, the role of the OD module is to surround each object with a 2D regressed bounding box (2 and 5 bounding boxes in images one and n, Fig. 1), which can be parameterized by its location [u, v], dimension [h, w], and orientation [θ, φ] relative to a predefined global frame (e.g., the one of the ego-vehicle which recorded the data). The shape and dimensions of the bounding boxes and the associated object's class can be used to estimate their 3D position and orientation, which are critical for safe driving. Such an estimation can be done using another CNN model [6]. While this technique is not as accurate as LIDAR for providing the depth information of objects, the estimation is enough for the AV to take proper actions. It correlates closer to human drivers' perception than LIDAR, which does not need accuracy under a few cm to take appropriate action.

### 2.2 Metrics used for Assessing Object Detection

#### 2.2.1 Object Detection Accuracy Metrics

The accuracy of an OD algorithm targeting AV applications refers to the level of similarity between the predicted and current bounding boxes, respectively, named the ground truth. The standard metric is the mean average precision (mAP) and the intersection over Union (IoU) metrics. IoU determines how much the predicted boundary box overlaps with the ground truth. Two other accuracy metrics are also occasionally used: Precision and Recall metrics. Precision refers to the percentage of correct predictions among all the predictions made. Recall is the ratio between the number of correct bounding boxes and the total number of target bounding boxes. Different applications may prioritize one over the other. For instance, it is critical not to miss any pedestrians in AV applications, so focusing more on the Recall metric is essential.

#### 2.2.2 Throughput and Latency Metrics

Besides accuracy, execution time is another critical metric. fps in OD is the throughput metric that measures the frequency at which images are processed. The latency represents the elapsed time between the acquisition of a current frame and the time when all the objects within that frame are detected. In AVs, the latency, which shall not exceed a few ms in all cases, depends on the object type and estimated depth. For instance, in Tesla's cars, front cameras with an FOV of 250 m can accommodate higher latency than one with a 60m FOV.

## 3. State Of the Art Algorithms Used In Autonomous Cars

Cameras-based OD algorithms used in AVs are cost-effective, and unlike LIDAR, they do not require complex postprocessing. Thus, they are favored by some car manufacturers and can be based on stereo cameras, multi-view (i.e., bird's eye view), or primarily monocular cameras.

### 3.1 Monocular-based CNN algorithms

All current commercial AVs use CNN to perform OD and classify monocular images. Traditionally, this task was handled using various mathematical and statistical approaches, such as the Hough Transform (HT), for detecting lanes within the roads (e.g., objects # 2 in Fig. 1) [7]. Other advanced OD techniques include Scale Invariant Feature Transform (SIFT) [8], Viola–Jones rectangles [9], and Haar-Like-wavelets, and Histogram of Oriented Gradient (HOG) [10]. This is because they require reasonable computation power in compact and low-power hardware platforms, which were available during that period. Indeed, their computation is like executing a single convolutional layer of CNN algorithms, which exhibits a local and highly parallel computation model. In addition, some of these algorithms, such as SIFT, are scale and rotation invariant and can accommodate partial occlusion, an essential requirement for OD in AV. For instance, in [11], an enhanced HOG algorithm that uses an additional radar sensor was used to detect and classify vehicles. In [12], HOG and SVM algorithms were used for vehicle detection during nighttime. However, they are limited to handling only a few real-life scenarios to detect relatively simple objects, compromising their practicability in AVs. Furthermore, they can detect only a single kind of object. Usually, several models are required to detect several models, for instance, one model for detecting vehicles [13], another one for detecting pedestrians [14], and another one for lane detection [15]. This makes their implementation to detect many classes complex. With the emergence of highly parallel SOC architecture, CNN-based models are exclusively used for OD on monocular images. They can cover many real-life scenarios and detect complex objects if adequately trained. One way to categorize these algorithms is to be single one-stage or two-stage detectors to emphasize accuracy and computation time, respectively.

#### 3.1.1 Two-stage object detection algorithms

This class of algorithms uses two sequential stages: a region (anchor) proposal generation stage, a classification, and a bounding box regression stage. While they are time-consuming, these algorithms usually yield better accuracy. Regions with CNN features (R-CNN) [16][17], Fast R-CNN, and to a greater extent, Faster R-CNN are usually adopted in this category. They all include a single network to accommodate the two phases sequentially to accelerate the execution time. In [16], the authors used R-CNN to achieve segmentation of an image into regions by a method they named 'selective search,' which merges nearby pixels that share certain similarities. They could produce final candidate regions by recursively combining the most similar regions into larger ones. This method was then used in [17] to extract up to 2,000 category-independent regions and feed them as candidate detections to their network called "Regions with CNN feature." This new method could achieve a high mAP of 53.7% on the PASCAL VOC 2010 dataset. This was slightly higher than Regionlets, which attained a mAP of 39.7%, using selective search region proposals. On Nvidia's K40 GPU, Fast R-CNN and R-CNN were executed in 2.3 and 49 seconds, respectively [17], which is still well below the real-time performance. The reason is that they both use time-consuming selective searches to find the region's proposals. This was tackled in the Faster R-CNN model, which consists of two modules (Fig. 2): The first deep, fully convolutional network module that determines candidate regions and the Fast R-CNN detector module that explores the proposed regions to classify eventual objects. The R-CNN uses an attention mechanism to reduce the search space.

Table 1 summarizes each R-CNN version's main characteristics and performance for AV applications [18]. While training the model with just the PASCAL VOC 2012 dataset had an accuracy of 67.0% mAP, training it with the combined set of PASCAL VOC 2007 and 2012 accomplished a 70.4% mAP. Also, when adding the MS COCO 2015 dataset to these two, the performance reached up to 75.9% mAP. The faster R-CNN model trained with the three datasets had mAP values for those categories relevant to autonomous driving: person, 84.1%; bike, 83.6%; bus, 81.9%; car, 82.0%; motorbike, 84.7%. Faster R-CNN achieved 36.2 mAP in the COCO 2016 dataset, running at 5 FPS on a NVIDIA M40 GPU.

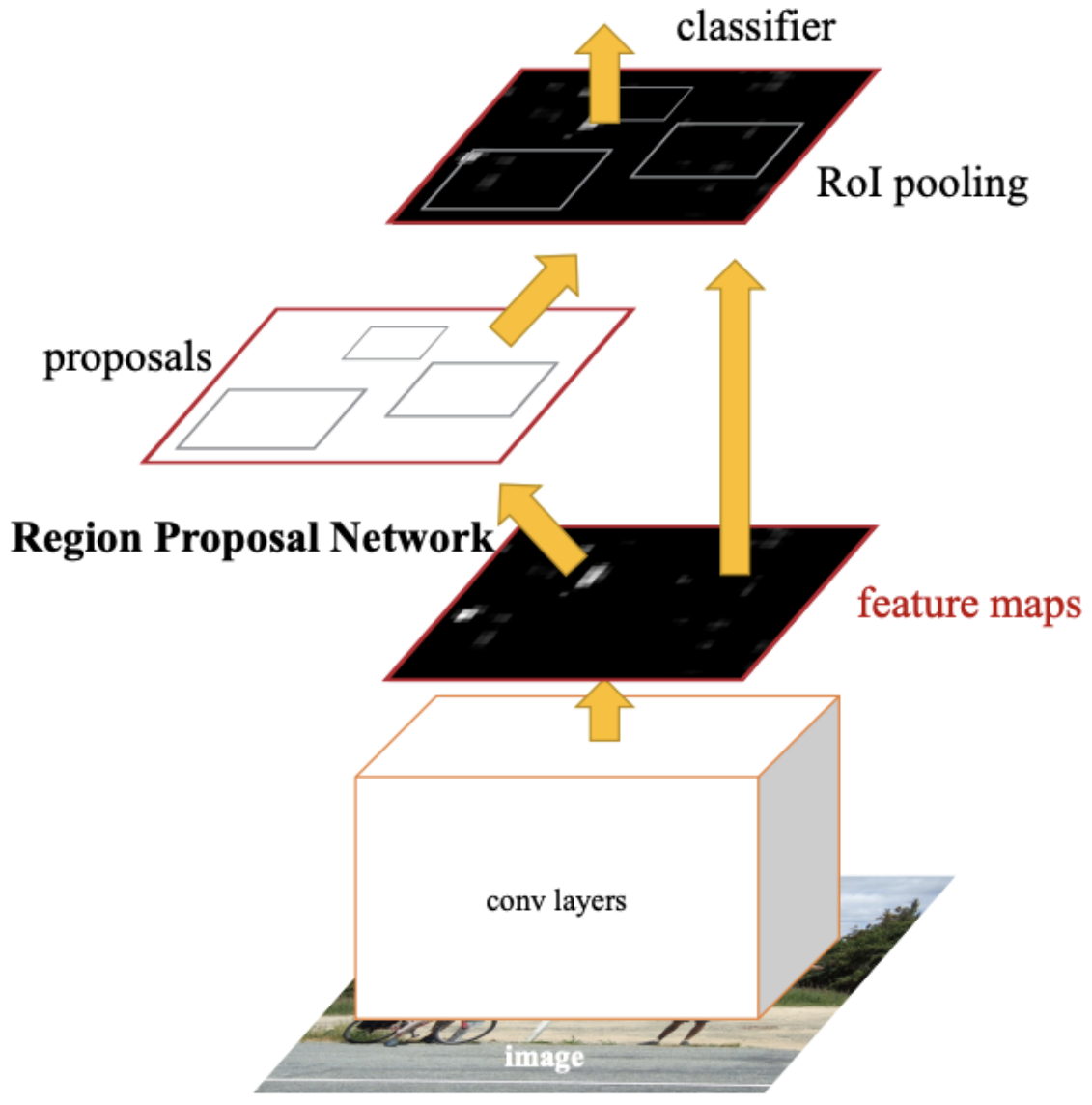

**Fig. 2**: Faster R-CNN structure

**Table 1**: Comparison between R-CNN, Fast R-CNN, and Faster R-CNN on an NVIDIA K40 GPU when trained with Pascal VOC dataset [18].

| | **R-CNN** | **Fast R-CNN** | **Faster R-CNN** |
|---|---|---|---|
| Region proposal method | Selective search | Selective search | Region proposal network |
| Processing time | 49s | 2.3s | 0.2s |
| mAP on PASCAL VOC 2007 | 58.5% | 66.9% | 69.9% |
| mAP on PASCAL VOC 2012 | 53.3% | 65.7% | 67.0% |

Also, in another work targeting AD application, [19] Faster R-CNN could achieve 94.4% accuracy in lousy weather conditions. Given these attributes and all its layers feature fine-grained parallelism, Faster R-CNN is likely one of the CNN models used in commercial AV.

### 3.1.2 One-stage object detection algorithms

In this category of algorithms, finding regions and classification tasks are combined into one step. In this case, the classifier is applied over a much larger set of candidate object locations, regularly sampled across the image, which densely covers spatial positions, scales, and aspect ratios. Extensive research done on this category has also significantly improved their accuracy, making them comparable to the ones corresponding to the double-stage methods. Yolo (you only look once) CNN model remains one of several car manufacturers' most accurate models for OD and classification likely used in AV [20]. The algorithm divides the input image into a grid, and each grid cell is taken as a bounding box to predict an eventual object. The algorithm determines a class probability and offset value for each bounding box. The boxes having a probability above a threshold value are selected and used to locate the object within that image; thus, multiple objects can be detected simultaneously. The latest Yolo v8 has four different classes (i.e., N, S, M. L, and X) depending on the scaling coefficient, where the various sizes of kernels and a different number of features are used (**Fig. 3**). The model is divided into the backbone for feature extraction, the neck, and the head modules. The neck module performs feature fusion to generate large, medium, and small-size features. It includes a layered C2f technique to improve deep image extraction and an SPPF layer to transform input feature maps into a fixed-size feature vector. This layer aims to enhance network responsiveness. The head module detects and classifies objects.

In [21], Yolo v5 was combined with the Generative Adversarial Network (GAN) model to achieve the highest accuracy of 96.75% for the night demonstrated. Nevertheless, one of the issues of Yolov8, and most other CNN models, including Faster CNN, is its accuracy degradation in the case of small OD, as it is challenging to detect image features, especially in complex road scenes. In [22], an enhanced Yolov8-QSD, a query-based method, enhances the algorithm's speed and accuracy for detecting small objects with an accuracy rate of 64.5% and a computation requirement of 7.1 GFOPS. However, this may still not be effective in practice. Commercial AVs, such as Tesla, use different FOV cameras to avoid missing any object on the road [1].

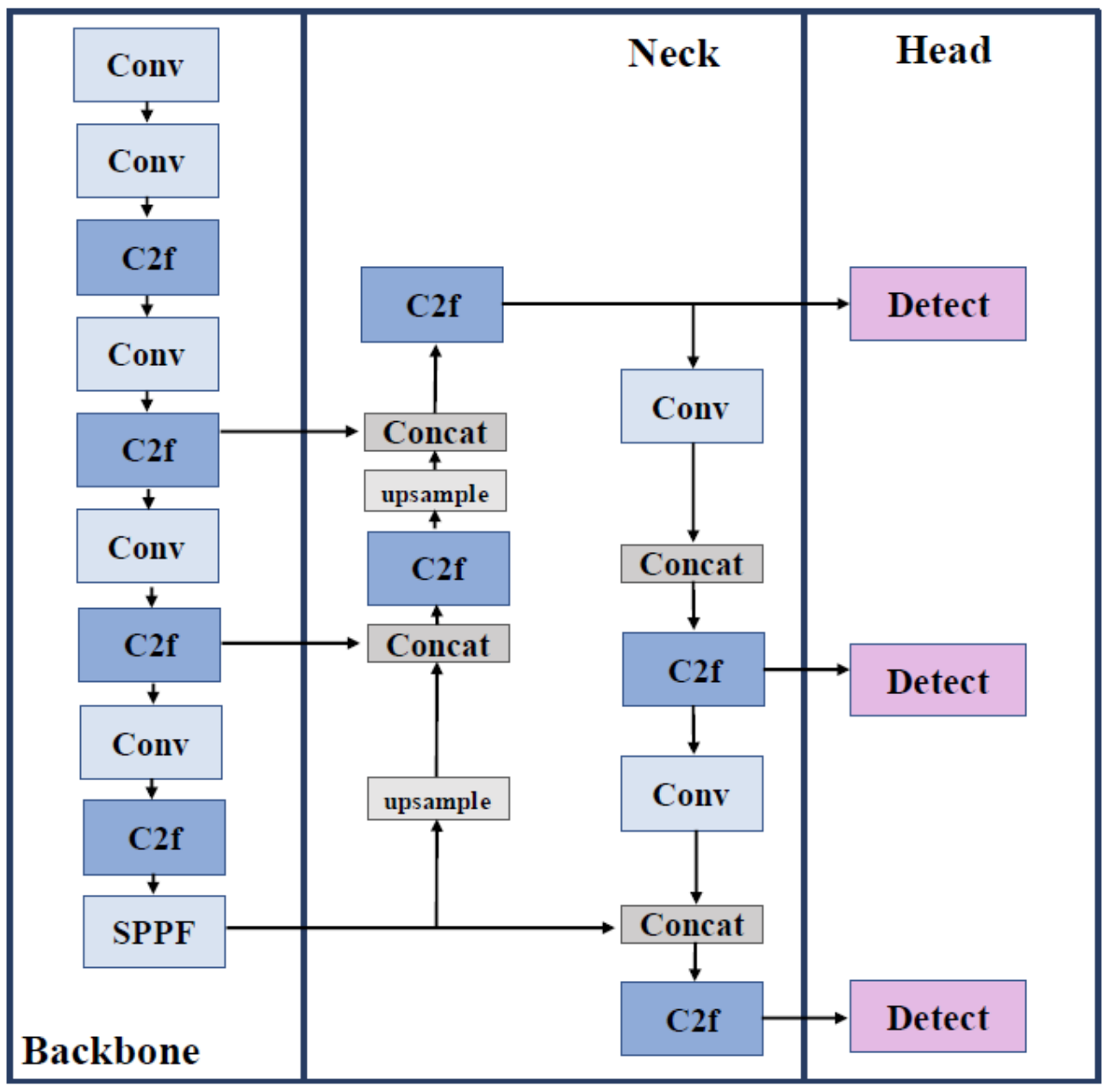

**Fig. 3:** Structure of Yolo v8

Single Shot Multibox Detector (SSD) was also widely used in AVs for OD. SSDs predicts object categories and bounding boxes directly from feature maps in a single forward pass. Unlike the YOLO model, the convolutional layers on SSD divide the input image into feature maps, i.e., grids of cells, of different sizes [23]. Each of the cells generates a fixed number of bounding boxes with different aspect ratios and scales, and a confidence score is given for every object class in all the boxes. This process is made in each feature map dimension, and those bounding boxes with high-class object scores are filtered. The main advantage of SSD over YOLO is that there is no feature resampling stage; instead, feature maps are used, allowing the network to be more accurate in detecting small objects and, simultaneously, making the whole process faster. On an Nvidia Titan and with the PASCAL VOC 2007 dataset, SSD achieves 74.3% mAP at 59 FPS, which outperforms Faster R-CNN (73.2% mAP at 7 FPS) and YOLOv1 (63.4% mAP at 45 FPS). In [24], the SSD algorithm was tested for various adverse conditions to achieve a mAp of 92.18% for a 15 ms/frame throughput. When tested on the KITTI database.

Although YOLO architecture differs across versions, most studies in the literature report broadly consistent training and tuning strategies. These configurations are critical for reproducing published results and for adapting YOLO to domain-specific tasks.

a) **Optimiser choice:** The majority of YOLO implementations rely on stochastic gradient descent (SGD) with momentum (typically 0.9–0.95) as the default optimizer, particularly for training from scratch [31]. More recent works, especially those employing transfer learning or fine-tuning with pretrained backbones, frequently adopt AdamW ($\beta_1 \approx 0.9$, $\beta_2 \approx 0.999$), which provides faster convergence and improved generalization in some scenarios [31].

b) **Learning rate scheduling:** Reported training commonly uses an initial warm-up phase (2–5 epochs) to stabilise gradients, followed by either cosine decay or one-cycle learning rate scheduling. The initial learning rate is scaled with the effective batch size (e.g., $lr_0 \approx 1e\text{-}3$ – 1e-2 for SGD; 1e-4 – 3e-4 for AdamW) [32], [33].

c) **Regularization parameters:** Across the surveyed papers, weight decay (L2 penalty) is consistently applied in the range of 1e-4 – 5e-4 to reduce overfitting [33]. Label smoothing (0.0–0.1) is occasionally reported, especially in datasets with noisy annotations [33]. Some YOLO variants (e.g., YOLOv5 and later) also incorporate Exponential Moving Average (EMA) of model weights, which helps stabilize evaluation.

d) **Loss functions:** Bounding-box regression typically employs GIoU, DIoU, or CIoU loss functions, which are improvements over traditional IoU loss. Classification and objectness branches generally use binary cross-entropy (BCE), with focal loss introduced in cases of high class imbalance.

e) **Data augmentation:** A strong emphasis is placed on data-level regularization. Common augmentations include mosaic augmentation, mixup, random scaling and translation, horizontal flipping, and HSV colour-space jittering. These augmentations are particularly valuable in autonomous driving datasets, where variability in lighting, scale, and occlusion is high.

f) **Other practices:** Many studies report the use of automatic mixed precision (AMP) to accelerate training and reduce memory consumption [32]. Gradient accumulation is occasionally applied when GPUs have limited memory. Early stopping, checkpointing based on validation mAP, and selection of the best validation model rather than the final epoch are frequently adopted strategies.

### 3.2 Stereo-camera-based object detection

CNN-based OD that relies on depth information was also used in some AVs as they may yield higher accuracy and provide depth information. They can also be used to determine objects' 3D position and velocity more accurately than the monocular technique. The availability of cost-effective, accurate, and real-time depth cameras significantly contributed to such integration. However, their accuracy is still far below that of LIDAR. However, in AV, and as mentioned recently by Tesla, depth information does not need to be accurate at cm precision. Depth-based 3D object detection remains underexplored, with only a few research efforts dedicated to this area, and none have been suggested for commercial AVs. Among research works from academia, one can cite the NMRF-stereo [25], Pseudo-label-based CNN [26], and stereo R-CNN [27]. This latest consists of performing object detection separately on a left and right image and then proceeding by matching the associated bounding boxes to estimate their depth. This doubles the computation requirement required for monocular-based depth computation and, thus, is unrealistic.

### 3.3 Object Detection algorithms in commercial AD systems and Pending Challenges

Tesla uses public and customized CNN models combined with machine vision and data fusion. It uses end-to-end training relying on a large dataset to consider the most significant number of scenarios. Yolo algorithm is likely one of the CNN models used as it has proven fast and accurate for AV. As a machine vision algorithm, real-time object tracking based on the Kalman filter is implemented to predict the motion of different obstacles. While this is required to avoid crashes, it adds a significant load to the processing unit as the tracking algorithms are time-consuming. Tesla's CNN model is dynamically reconfigured based on new updates.

In contrast to Tesla's autopilot, which relies only on cameras, Waymo's AV uses a combination of cameras, Lidar, and radar sensors. In addition to object detection (both 2D and 3D object detection using LIDAR), semantic segmentation and object tracking are also implemented. Thus, the Waymo car tends to build a 3D environment for vehicles to enhance its safety. 2D detection is based on Yolo v2 to detect objects like cars, pedestrians, traffic lights, and road signs. The object tracking is based on Faster R-CNN to track objects in the point cloud data from LIDAR sensors. It also uses SqueezeSeg, a customized CNN architecture, to detect objects from point cloud data. As mentioned earlier, none of the AV vendors explained how they implemented OD for multiple cameras.

Academic researchers have suggested many models and improvements to CNN models for AVs. For instance, in [28], the authors have considered the practical challenges that AVs may face, such as fog, low light, and sun glare, for detecting vehicles using three CNN models, namely ResNet, GoogleNet, and SqueezeNet, with transfer learning. ResNet achieved an accuracy of 100%, while GoogleNet and SqueezeNet achieved 65.50% and 90.0% accuracy, respectively. Other CNN models include DDCDC [29], Obmo [30], and shape aware [15]. Nevertheless, as illustrated in **Table 2**, most of the research works were solely done on one kind of object at most. Assessing their performance for multiple objects is rarely tackled in academia. According to [16] concerning AV, 40 %, 32 %, and 28% of academic research was conducted to detect vehicles, pedestrians, and lanes, respectively. Most of them suggested CNN-based algorithms, especially after 2008.

**Table 2:** CNN Models used in AD for different objects.

| **Object to be detected** | **Year/ Reference** | **CNN model** | **Accuracy** | **Note** |
|---|---|---|---|---|
| Cars | 2022 [11] | SqueezeNet, ResNet-50, EfficientNet-b0 | 98.48%, 97.78%, 96.25% | Tested on cloudy, rainy, snowy, sandy, shiny, and sunrise |
| | 2022 [11] | YOLO-v5 | 96.75% | Tested during night |
| Pedestrians | 2022 [12] | YOLO-v3 | 74.38% | Tested under several weather conditions |
| | 2023 [6] | YOLO-v7 | mA=0.73 | Rain, Fog, Snow, Sand |
| Lane detection | 2022 [15] | Resnet18 | 97% | Tested on rain conditions. Achieved 99% accuracy in good weather. |
| | 2022 [28] | Modified YOLO-v3 | 90.27% | Tested in KIT, Toyota Technological Institute, and Euro Truck Simulator 2 |

Besides single- and double-stage OD algorithms, attention-based OD has also gained interest in AD applications in recent years. Inspired by how humans concentrate on relevant information, the main idea of these algorithms was the concept of significance, which consists of pushing the model to focus on portions with a significant amount of valuable information. This was achieved by adding attention weights representing the relative importance of each element concerning each of the other components in a sentence [9]. An attention-based model called a transformer, which comprises encoder-decoder layers, was proposed in [10]. Hence, a multi-head attention encoder consisting of several attention layers running in parallel was proposed, with the output stacked and fed to the decoder. Building on this concept, the Detection Transformer (DETR) [19] reformulates object detection as a direct set-prediction task, eliminating handcrafted components such as anchor boxes or non-maximum suppression. The architecture of DETR is illustrated in **Fig. 4**. DETR begins with a CNN backbone (e.g., ResNet-50) that extracts image features, which are then flattened and combined with positional encodings to preserve spatial structure. These features are passed into a transformer encoder, where multi-head self-attention computes global interactions using the standard query–key–value mechanism: each feature vector is projected into queries (Q), keys (K), and values (V), and their attention weights highlight relevant contextual relationships across the image.

A transformer decoder then operates on a fixed set of learnable object queries, which act as placeholders for potential objects. Through cross-attention, the decoder aligns these queries (Q) with the encoded image features (K,V), allowing each query to attend selectively to specific regions of interest. The refined embeddings are passed to simple feed-forward networks that output both a class label (including a "no-object" category) and normalized bounding box coordinates. A Hungarian bipartite matching loss enforces a one-to-one correspondence between predicted and ground-truth objects, ensuring that each query represents a unique detection. This end-to-end design enables DETR to recognize individual instances, even when multiple objects belong to the same class, while leveraging global context. In [2]9], it was demonstrated that DETR achieves a higher mAP accuracy and throughput with only 41 M parameters than the conventional CNN model with 166 M parameters.

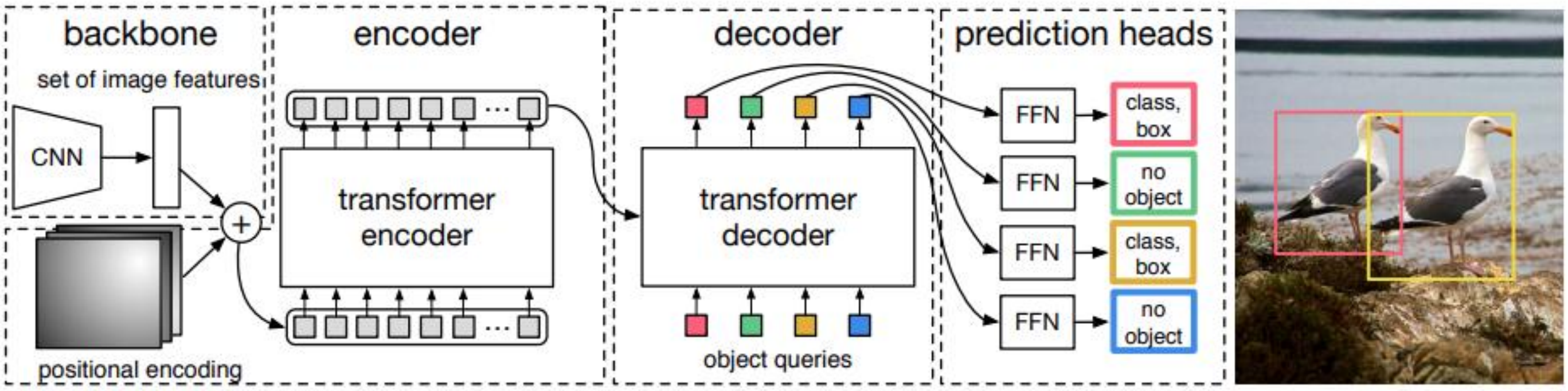

**Fig. 4:** Architecture of DETR [38].

**4. Hardware Accelerators Used in Autonomous Cars**

Historically, multicore CPUs, GPU, FPGAs, and ASICs were the leading platforms for real-time OD. This was motivated by the highly fine-grain parallelism of most OD algorithms, including AI and non-AI algorithms. They typically comprise many processing elements (PEs), interconnected in mesh via dedicated links, each with a relatively small local data memory. Usually, each PE is connected to 2 or 4 other PEs. All of them are issued with a standard instruction bus to form a Single Instruction, Multiple Data (SIMD) architecture, a parallel processing model where the same instruction is applied simultaneously across multiple data points (e.g., pixels or matrix elements), which is especially effective for image and neural network computations. This architecture is easily implemented into ASIC, FPGA, or GPU cores because of its regularity. Hence, the processing is done locally on neighboring pixels and can be conducted parallel throughout the image space. Significant progress in the semiconductor industry has led to the emergence of system-on-chip (SOC) chips, which simultaneously include at least two different technologies. Such heterogeneous architecture is anticipated as the computation models of associated algorithms used in AVs, exclusively AI-based in commercial AV systems, are highly intensive. Hybrid SOC technology is used in the HAs of the vision systems of all AV manufacturers. For instance, all HAs incorporated an ASIC module for decoding multiple high-definition video channels to seamlessly interface with video cameras, which mainly generate standard H265/H264 video standards [1]. Another notable ASIC module available in most AVs' OD modules is the one that performs highly parallel matrix multiplications, as this is required for different layers of AI models (i.e., convolution layers and max pool layers) and is the most computationally time-consuming task.

It is worth noting the evident gap of most academic research works for not considering the real-field constraints of AVs. For instance, most object detection algorithms were mapped onto cloud GPUs, such as 3090 [12], which is power angry and features an optimized form factor. In addition, most of the work seeks to detect fewer objects than what may find on the roads. In addition, they consider only one camera channel and rely on datasets that are not comprehensive enough.

### 4.1 Tesla's HW4 Hardware Accelerator

Tesla's HW4 hardware board (Fig. 5a) can support up to 12 cameras, even though the current Tesla autopilot system comprises only eight cameras. Based on the Linux kernel, the system consists of two (2) 7 nm FSD-2 (full-scale driving) customized processors based on Samsung's Exynos-IP core. Each processor comprises 20 CPU cores (5 clusters of 4 ARM Cortex A72 cores each), each operating at 2.35 GHz, and three network processing units (NPUs), each operating at 2.2 GHz, to deliver an aggregated 121.65 TOPS of performance. In each cycle, 256 activation and 128 weight data are read from the SRAM to the Multiply Accumulate units (MACs). MACs are specialized circuits that perform multiplication followed by addition in a single step and form the fundamental arithmetic building blocks of neural network accelerators. The MACs design is a grid, with each NPU core having a 96×96 grid for 9,216 MACs and 18,432 operations per clock cycle. The board consumes around 16 V for 10 A (160 W power consumption), which requires different cooling pipes. [1]. The board has a memory capacity of 16 GB RAM and 256 GB of storage, an improvement over the previous version, HW-3, which can't be retrofitted to existing cars in a simple board swap. The most recent version, v13.2, released in November 2024 and available in Tesla's cybertracks, is thought to yield full AV navigation.

At the architectural level, Tesla FSD NPU [34] depicted Fig. 5b-c provides block-level confirmation of its internal organization. The image shows a large central MAC array surrounded by multiple on-die SRAM banks and smaller blocks for activations, pooling, and write-back. This organization is typical of modern NPUs and explains how HW4 sustains convolution-heavy workloads such as object detection. The dense MAC array executes the arithmetic required by convolutional layers, while the nearby SRAM banks and activation buffers enable on-chip data reuse, reducing external memory transfers and improving throughput. The smaller activation and pooling units integrated next to the compute array minimize the need for intermediate off-chip operations, lowering latency and energy cost. Such a design is particularly advantageous for one-stage detectors like YOLO, which demand high parallelism and efficient memory utilization.

An upgraded HW5 version, expected to be released in January 2026, outperforms the HW4 version 10 times [1][2] with up to 800 W power consumption. It is aimed at being deployed into next-generation level 5 cars. Tesla started delivering new Tesla vehicles, including models S, X, and Y, with the HW4 accelerator in early 2023. Nevertheless, some features are not yet operational on all models, particularly Cybertruck.

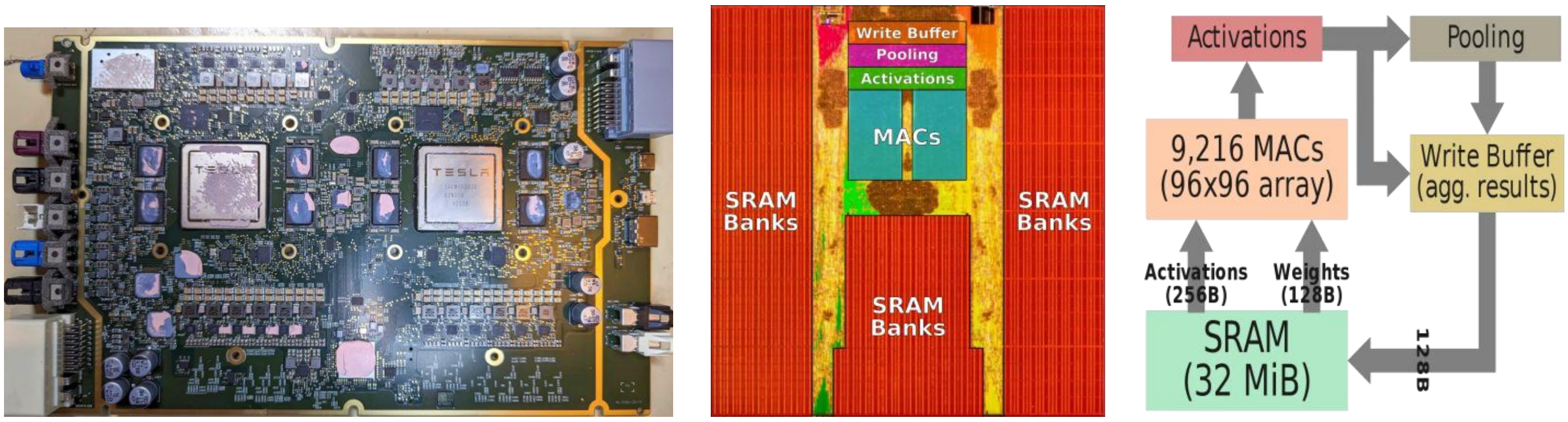

(a) (b) (c)

**Fig. 5:** (a) Tesla's HW4 Hardware board (b) Tesla FSD NPU die [34] (c) Tesla FSD chip NPU block diagram [34].

### 4.2 Nvidia's AGX Drive GPU and other Processors

NVIDIA provides a series of edge GPUs, which have been demonstrated to be effective for AV applications. Like Tesla's HW4 processor, NVIDIA's edge GPUs include ASIC modules for real-time video decoding and a highly parallel ASIC module, Deep Learning Accelerator (DLA), for power-effective execution of various layers required in CNN models, such as convolution layers. It also comprises a multicore CPU, mainly dedicated to high-level control operations. However, it differs from Tesla's processors by having an integrated massive parallel GPU core that includes several stream multiprocessors (SMs) sharing L2 memory cache and that can directly access the main SDRAM memory. This GPU core complements the DLAs by accelerating a broader range of parallel workloads, including both convolutional and transformer-based detection tasks.

The flagship DRIVE AGX Orin platform exemplifies this heterogeneous architecture. Built on the Orin-X SoC, it combines 12 ARM Cortex-A78AE CPUs, the Ampere Tensor Core GPU, and specialized accelerators including two second-generation DLAs, a Programmable Vision Accelerator (PVA) for classical computer vision functions, and an Optical Flow Accelerator (OFA) for motion estimation and tracking. In total, the system delivers up to 254 INT8 TOPS with 205 GB/s memory bandwidth, supported by 32 GB LPDDR5 RAM. The use of on-chip SRAM buffers within each accelerator reduces external memory traffic and latency, enabling real-time execution of object detection pipelines such as YOLO across multiple camera feeds.

The DLA is a hardware engine purpose-built for efficient inference of convolutional neural networks. NVIDIA's Orin platform employs the second-generation NVDLA 2.0 design, offering up to nine times the performance of its predecessor. Structurally, DLA 2.0 includes a small microcontroller and a hardware scheduler that orchestrates operations as illustrated in Fig. 6. It features dual convolution units, each paired with a dedicated 608 KB local buffer, which is used during convolution processing as well as for spatial data reuse. These buffers are fed from a larger 1 MB shared buffer, serving as a multi-layer staging area within the accelerator. After convolution, data is passed to post-processing blocks, which handle operations such as activation functions and pooling. The accelerator also supports depthwise convolutions, and includes optimizations for structured sparsity, enabling pruning-aware inference for speed and energy gains. The DLAs deliver a total of 87 INT8 TOP/s [35].

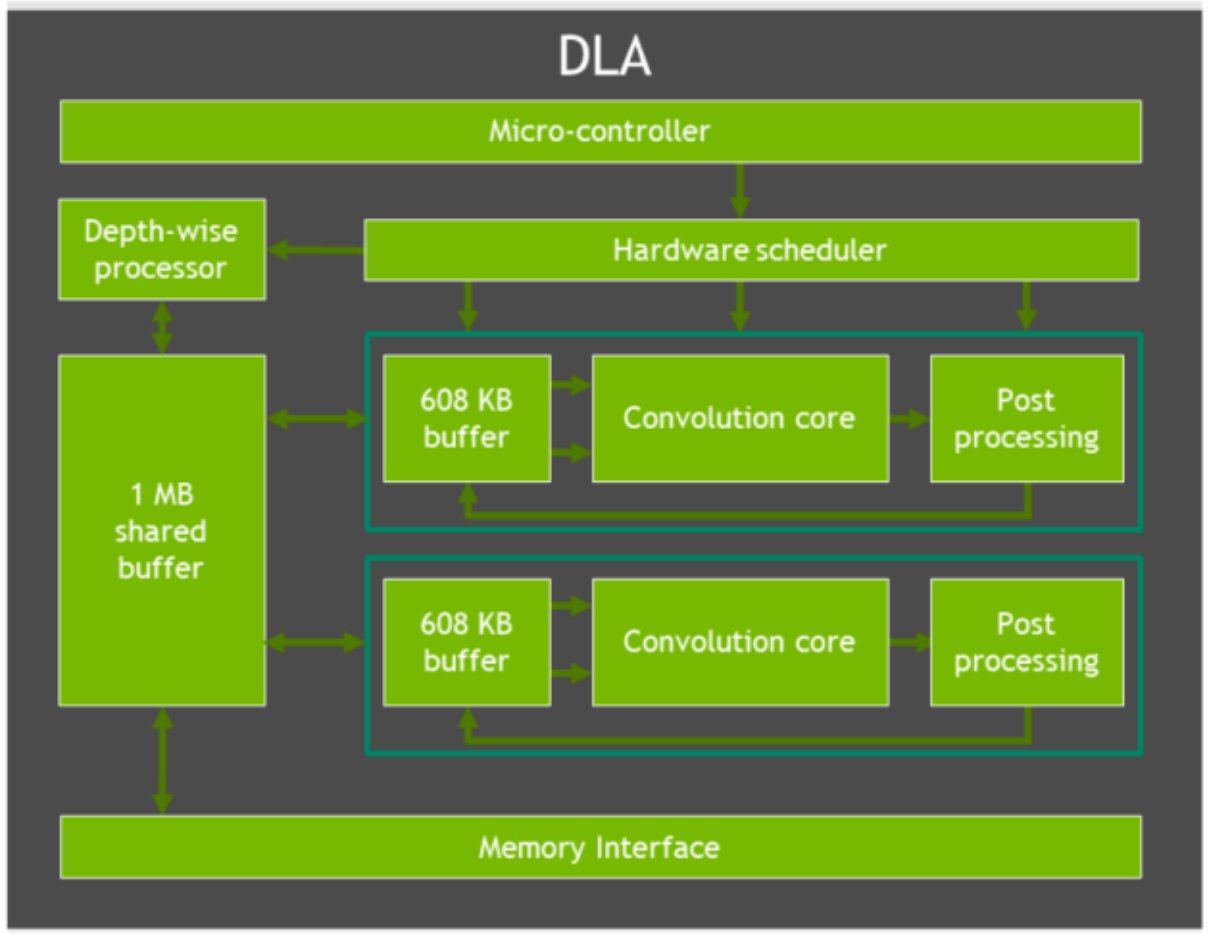

**Fig. 6:** Architecture of DLA v2 [36].

Complementing the DLA, the PVA uses two Vector Processing Subsystems (VPS) as shown in Fig. 7. Each VPS includes a Vector Processing Unit (VPU), an instruction cache, and local vector memory (VMEM), with a dedicated Decoupled Look-Up Table (DLUT) for efficient parallel table lookups without memory bank conflicts. The two VPS blocks share a common L2 SRAM, providing fast on-chip storage for intermediate data. Data movement is managed by dual hardware DMA engines, each supporting five-dimensional addressing and up to 16 hardware channels, capable of sustained transfer rates between 10–15 GB/s depending on workload intensity. These are coordinated by an embedded ARM Cortex-R5 controller, which handles scheduling and synchronization. In practice, the PVA efficiently accelerates feature extraction, image filtering, stereo vision, and optical flow estimation, offloading these stages from the GPU and DLAs The PVA delivers up to 2048 INT8 GMAC/s [35]. By reducing GPU load and minimizing external memory traffic, the PVA enables more balanced execution of end-to-end detection pipelines, where CNN inference runs on the GPU/DLA while pre- and post-processing is handled at much lower energy cost.

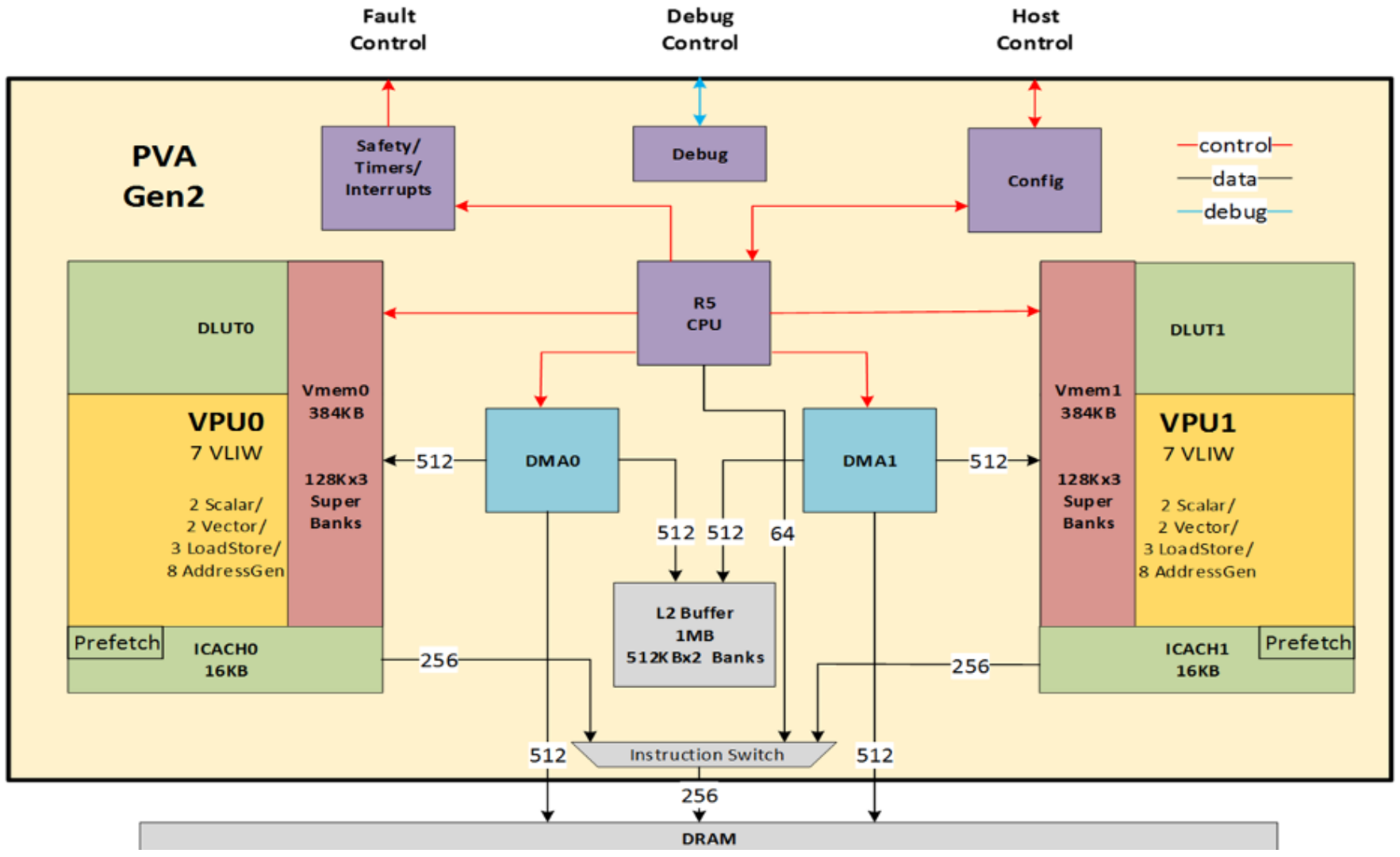

**Fig. 7:** Architecture of PVA Gen2 [37].

From an automotive perspective, DRIVE AGX Orin is further distinguished by its safety features. It integrates lock-step R52 safety islands and is compliant with ISO 26262 standards (ASIL-B at chip level, ASIL-D systematic). Running the DRIVE OS software stack, which includes CUDA, TensorRT, and DriveWorks middleware, the platform provides a safety-certified environment with deterministic scheduling for perception workloads. Rich automotive I/O, including GMSL camera interfaces, high-speed Ethernet, CAN, and FlexRay, ensures seamless sensor fusion from cameras, LiDARs, and radars. These combined features make NVIDIA's Orin-based edge GPUs not only versatile but also purpose-built for the real-time, safety-critical object detection workloads of autonomous vehicles.

Table 3 enlists some recent NVIDIA edge GPU modules that can be potentially used in AVs [3]. For comparison purposes, it also lists some of the recent cloud GPUs. Fig. 8 compares the NVIDIA edge and cloud GPUs based on GPU cores, maximum frequency, speed and power by normalizing the values in Table 3. Among these, cloud GPUs are the fastest, with many cores than edge GPUs by trading off power. Among the edge GPUs, AGX Orin, with high speed, frequency and cores, is suitable for autonomous cars. Jetson TX1 is adopted for low power applications.

**Table 3:** NVIDIA GPU modules (edge + cloud) specifications [3].

| Jetson Module | GPU cores | Max frequency [MHz] | Speed [GFLOPs] | Power [W] |
|---|---|---|---|---|
| **Edge GPU** | | | | |
| Jetson nano | 128 | 921 | 472 | 5-10 |
| Jetson TX1 | 256 | 998 | 1022 | <6 |
| Jetson TX2 | 256 | 1300 | 1330 | 7.4-15 |
| Jestson Xavier NX | 384 | 1100 | 2100 | 10-20 |
| Jetson AGX Xavier | 512 | 1377 | 3200 | 10-30 |
| AGX Orin | 2048 | 1300 | 10649 | 60 |
| **Cloud GPU** | | | | |
| Tesla T4 | 2560 | 585 | 8100 | <700 |
| Quadro RTX 5000 | 3072 | 1815 | 112000 | <230 |
| GTX Titan X | 3072 | 1000 | 6691 | <250 |
| Titan RTX | 4608 | 1350-1750 | 130000 | <280 |
| A100 | 6912 | 1095 | 312000 | <250 |

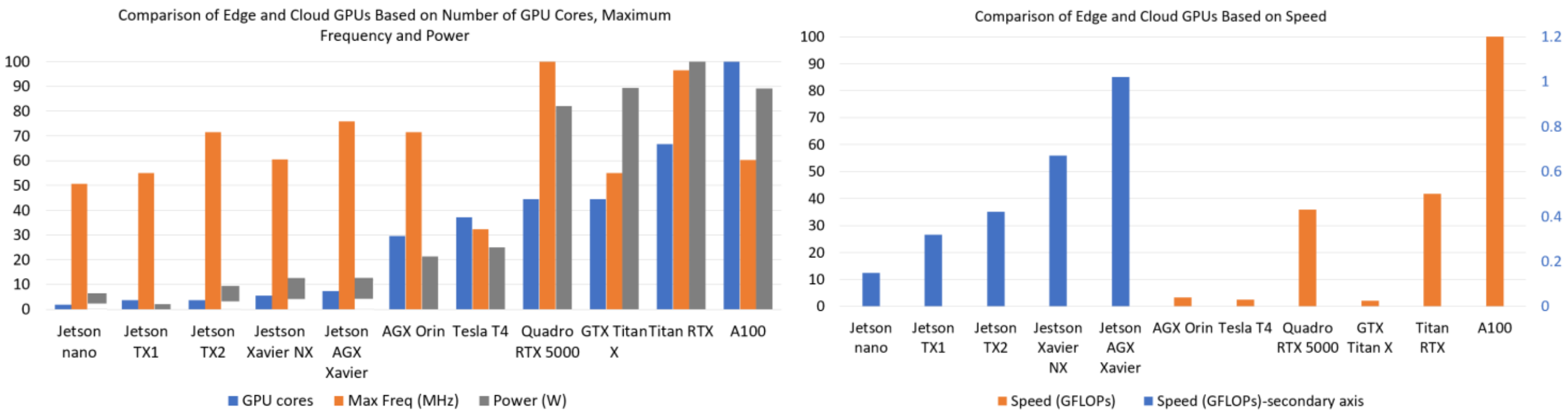

(a) (b)

**Fig. 8:** Comparison of NVIDIA edge and cloud GPUs based on (a) GPU cores, maximum frequency and power (b) speed. The series representing blue colored matric is plotted on a secondary y-axis (right) to allow simultaneous visualization of both small (0.5–0.6) and large (up to 100) values. The values from Table 3 are normalized for visualization.

Table 4 summarizes solutions deployed on conventional cloud GPU, whereas Table 5 enlists object detection solutions for edge GPUs. To complement the detailed numerical comparison in Table 5 and Table 6, we generated bubble charts (Fig. 9 and Fig. 10) to highlight key trade-offs across hardware–model implementations. Each bubble represents a platform-model pair. The horizontal axis corresponds to throughput (FPS), the vertical axis to accuracy (mAP), while the bubble size encodes a derived computational efficiency metric. Since most reviewed works did not report actual power consumption in Watts, direct energy efficiency (fps/W) could not be evaluated. Instead, we introduce a computational efficiency metric, defined as the ratio of throughput (FPS) to the computational cost (BFLOPs per frame):

$$\text{Computational Efficiency} = \frac{\text{FPS}}{\text{BFLOPs}}$$

This reflects how effectively a hardware–model pair converts raw computation into useful throughput. Larger values indicate more efficient utilization of compute resources.

The visualizations reveal several trends. Models such as Tiny YOLO variants achieve relatively high FPS but often at the cost of lower accuracy. In contrast, heavier architectures such as Faster R-CNN provide higher accuracy but yield much lower throughput and computational efficiency, making them less suitable for real-time AV deployment. Among cloud GPUs, GTX Titan X and NVIDIA A100 gives worst accuracy and performance and are computationally inefficient. Within the same platform, differences across models are clearly visible: for example, Jetson Nano favors lightweight YOLO derivatives, whereas higher-end GPUs (e.g., NVIDIA A100, Quadro 5000) can sustain both higher accuracy and larger model complexity while still achieving competitive efficiency. The most favorable trade-offs are observed with one-stage detectors such as YOLOv4, YOLOv5, YOLO-ReT, TRC-YOLO and more recent transformer-based detectors such as DINO, particularly when deployed on mid- to high-end accelerators (e.g., NVIDIA Xavier NX, AGX Xavier, Tesla T4, and Quadro 5000). These combinations achieve both real-time throughput (>30 fps) and robust accuracy (>70% mAP) while maintaining reasonable computational efficiency. Such characteristics make them the most viable candidates for deployment in commercial AV platforms, where a balance between speed, accuracy, and efficient hardware utilization is essential.

Recently, Nvidia plans to deliver DRIVE Thor with over 2,000 TeraFLOPS computation power to automotive customers, more than 15 times what Tesla's new HW4 can provide. Based on the data in Table 4 where 134.44 and 10.3 BFLOPS are required for Faster RCNN and Yolo v4, such platform is expected to easily accommodate more than 10 cameras simultaneously.

**Table 4:** Object detection systems in conventional cloud GPU.

| Platform | Model | Dataset | Accuracy [mAP%] | Performance [FPS] | Processing Power Needed [BFLOPs] | Size [MB] | Ref. |
|---|---|---|---|---|---|---|---|
| Tesla T4 | Faster R-CNN | Vehicle Aerial Imaging from Drone (VAID) 512×512 images | 90.2 | 25 | 134.4 | 330 | [20] |
| | YOLO v2 | | 87.6 | 67.4 | 44.45 | 202 | |
| | Tiny YOLO v2 | | 72.7 | 191.6 | 8.1 | 44.2 | |
| | YOLO v3 | | 96.5 | 34.7 | 99 | 246 | |
| | Tiny YOLO v3 | | 87 | 191.2 | 8.25 | 34.8 | |
| | YOLO v4 | | 97.1 | 34.2 | 90.29 | 256 | |
| | Tiny YOLO v4 | | 93.7 | 71.5 | 10.3 | 23.6 | |
| | YOLO-RTUAV | | 94 | 176.4 | 7.77 | 5.8 | |
| Quadro 5000 | Faster R-CNN | VAID 512×512 images | 90.2 | 17.18 | 134.4 | 330 | [20] |
| | YOLO v2 | | 87.6 | 92.2 | 44.45 | 202 | |
| | Tiny YOLO v2 | | 72.7 | 274.7 | 8.1 | 44.2 | |
| | YOLO v3 | | 96.5 | 43.0 | 99 | 246 | |
| | Tiny YOLO v3 | | 87 | 274.0 | 8.27 | 34.8 | |
| | YOLO v4 | | 97.1 | 37.8 | 90.29 | 256 | |
| | Tiny YOLO v4 | | 93.7 | 305.8 | 10.3 | 23.6 | |
| | YOLO-RTUAV | | 94 | 214.1 | 7.77 | 5.8 | |
| GTX Titan X | YOLO v4 | MS COCO 512×512 images | 43 | 31 | 52 | - | [23] |
| | LRF | | 37.3 | 31.3 | - | - | [24] |
| | RFB-Net | | 33.8 | 33.3 | - | - | [25] |
| | SSD | | 76.8 | 22 | - | - | [9] |
| | RefineDet | | 41.8 | 54.1 | - | - | [26] |
| | M2det | | 37.6 | 18 | - | - | [27] |
| | PFP-NetR | | 35.2 | 24 | - | - | [28] |
| | CornerNet | | 40.5 | 4.4 | - | - | [29] |
| A100 NVIDIA | DINO | MS COCO 416×416 images | 19 | 24 | 279 | - | [30] |
| | Faster R-CNN | | 37.9 | 21 | 207 | - | |
| | DETR | | 15.5 | 20 | 225 | - | |
| | HTC | | 42.3 | 5 | 441 | - | |

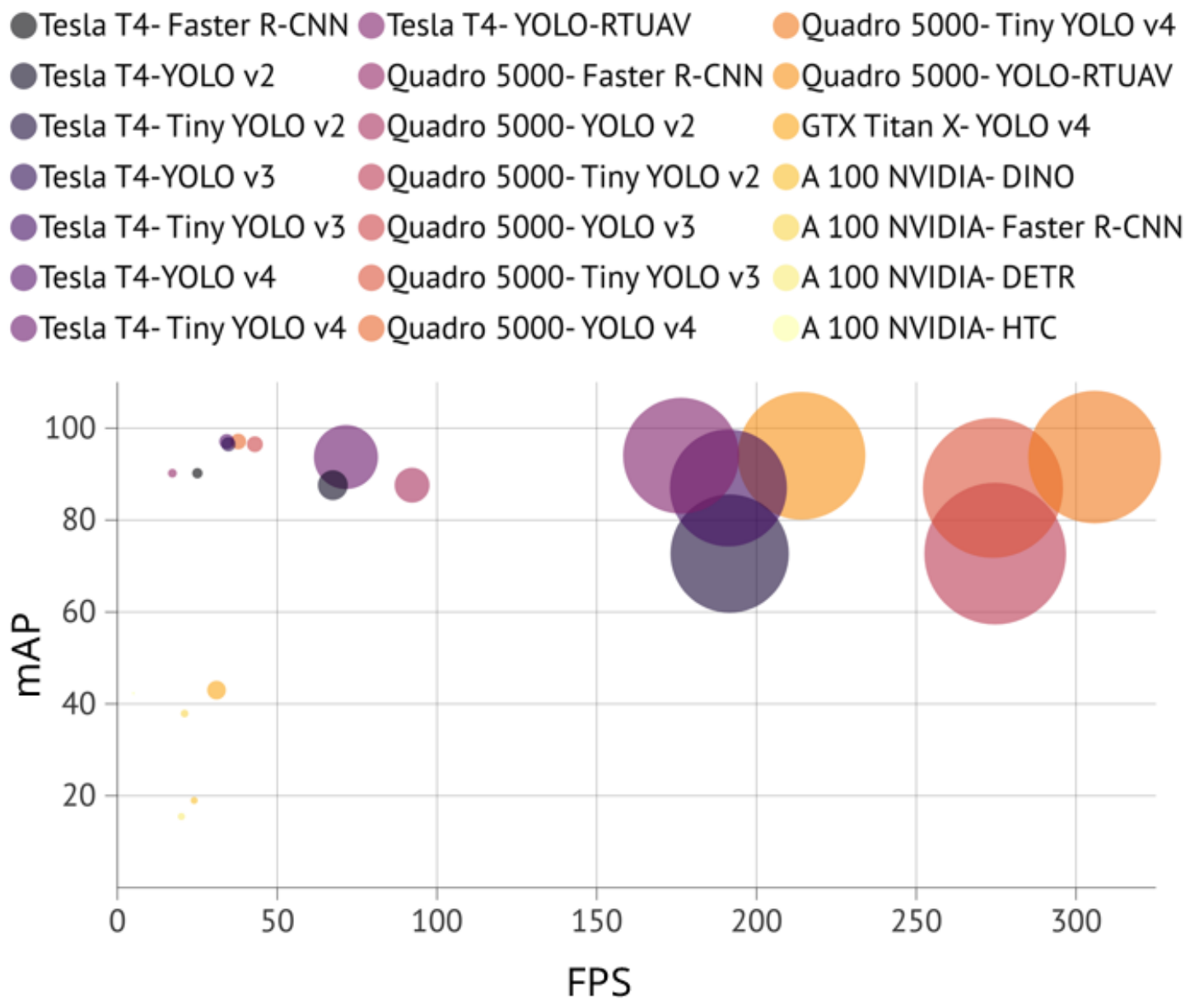

**Fig. 9**: Comparison of cloud GPUs based on accuracy, performance and computational efficiency.

**Table 5**: Comprehensive table for object detection in edge GPU

| Platform | Model | Dataset | Accuracy [mAP%] | Performance [FPS] | Processing Power Needed [BFLOPs] | Size [MB] | Ref. |
|---|---|---|---|---|---|---|---|
| Jetson AGX Xavier | Tiny YOLO v2 | PASCAL VOC 416×416 images | 57.1 | - | 6.97 | 60.5 | [14] |
| | Tiny YOLO v3 | | 58.4 | - | 5.52 | 33.4 | |
| | YOLO Nano | | 69.1 | 48.2 | 4.57 | 4 | |
| | MobileNet-SSD | PASCAL VOC 512×512 images | 72.4 | 17 | - | 25 | [15] |
| | MobileNet-SSD v2 | | 75.6 | 21 | - | 32 | |
| | YOLO-ReT | PASCAL VOC 320×320 images | 68.75 | 95.97 | - | 5.2 | [16] |
| Jetson Xavier NX | YOLO v2 | PASCAL VOC 416×416 images | 63.5 | 6.9 | 29.37 | 202.7 | [17] |
| | Tiny YOLO v2 | | 44.53 | 21.2 | 6.98 | 63.5 | |
| | Tiny YOLO v3 | | 52.78 | 30.5 | 5.45 | 34.9 | |
| | Tiny YOLO v4 | | 62.57 | 29.2 | 6.82 | 23.7 | |
| | TRC YOLO | | 66.4 | 36.9 | 2.98 | 17.8 | |
| | SqueezeNet SSD | PASCAL VOC 416×416 images | 64.3 | 11.4 | 4.50 | 22.7 | [18] |
| | MobileNet SSD | | 66.1 | 16.2 | 4.29 | 63.5 | |
| | Trident-YOLO | | 67.6 | 29.3 | 5.2 | 10.9 | |
| | | MS COCO 416×416 images | 40.3 | 29.3 | - | 416 | |
| | Tiny YOLO v3 | | 283 | 26.2 | - | 416 | |
| | Tiny YOLO v4 | | 35.8 | 25.3 | - | 416 | |
| | YOLO-ReT | PASCAL VOC 320×320 images | 68.75 | 71.64 | - | 5.2 | [16] |
| Jetson Nano | YOLO-ReT | PASCAL VOC 320×320 images | 68.75 | 33.19 | - | 5.2 | [16] |
| | | MS COCO 320×320 images | 34.91 | 33.19 | - | 5.2 | |
| | Tiny YOLO v3 | MS COCO 416×416 images | 33.1 | 17 | 5.57 | 996 | [19] |
| | Tiny YOLO v4 | | 40.2 | 18 | 6.91 | 1018 | |
| | SF-YOLO | | 37.7 | 26 | 3.69 | 926 | |
| | Faster R-CNN | Vehicle Aerial Imaging from Drone (VAID) 512×512 images | 90.2 | 1.3 | 134.4 | 330 | [20] |
| | YOLO v2 | | 87.64 | 2.4 | 44.45 | 202 | |
| | Tiny YOLO v2 | | 72.7 | 41 | 8.1 | 44.2 | |
| | YOLO v3 | | 96.5 | 1.4 | 99 | 246 | |
| | Tiny YOLO v3 | | 87 | 5.0 | 8.27 | 34.8 | |
| | YOLO v4 | | 97.1 | 1.2 | 90.3 | 256 | |
| | Tiny YOLO v4 | | 93.7 | 12.4 | 10.3 | 23.6 | |
| | YOLO-RTUAV | | 93.7 | 12.4 | 10.3 | 23.6 | |
| Jetson TX1 | Tiny YOLO v3 | PASCAL VOC | 61.3 | 21.6 | 5.47 | 34.9 | [21] |
| | Tinier YOLO | | 65.7 | 25.1 | 2.56 | 8.9 | |
| Jetson TX2 | EfficientDet-Lite | KITTI 512×512 images | 76.1 | 16.5 | 2.2 | 3.2 | [22] |
| | Tiny YOLO v3 | | 73.9 | 37.4 | 8.3 | 8.7 | |
| | | | 31.1 | 45 | 5.57 | 1044 | [19] |
| | Tiny YOLO v4 | MS COCO 416×416 images | 40.2 | 44 | 6.91 | 1053 | |
| | SF-YOLO | | 33.7 | 62 | 3.69 | 960 | |
| Jetson RTX | Tiny YOLO v3 | 416×416 images | 31.1 | 166 | 5.57 | - | [19] |
| | MS COCO | | 40.2 | 165 | 6.91 | - | |
| | SF-YOLO | | 33.7 | 168 | 3.69 | - | |

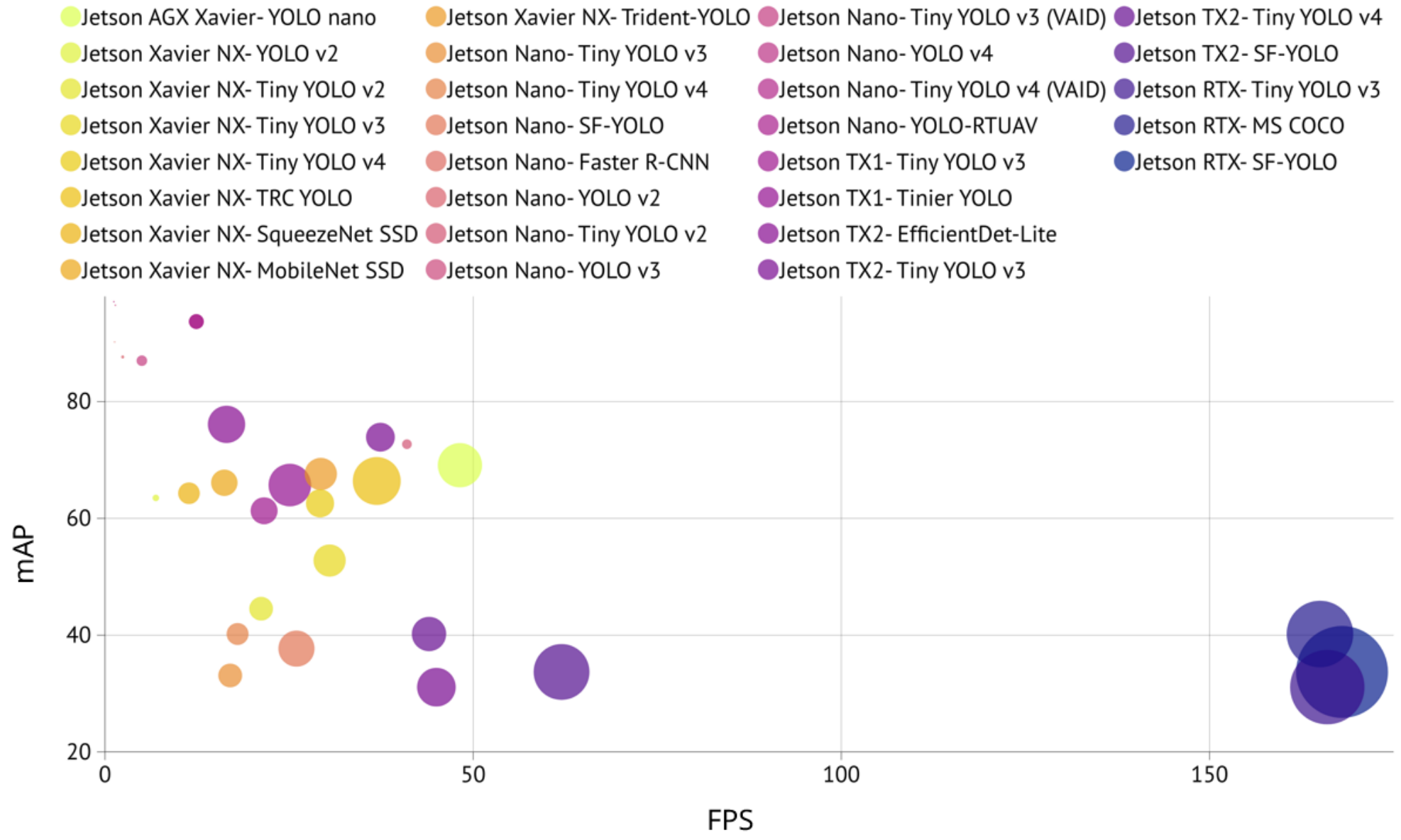

**Fig. 10:** Comparison of edge GPUs based on accuracy, performance and computational efficiency.

### 4.3 Other Hardware Architectures for Object Detection Targeting Autonomous Vehicles and Discussions

Alphabet Waymo and GM Cruise use full-sized GPUs that cost 10x more in their cars during development and early testing and are looking to make their own much faster (and more expensive) SoCs for their vehicles. Other edge hardware architectures that combine multicore CPU, programmable logic, and ASIC on a single chip, such as Xilinx's Versal FPGA architecture, have recently attracted the attention of researchers to reduce even further the power consumption of GPUs while offering reasonable computation power [2]. This technology can be a good alternative to GPU, provided that more user-friendly hardware-software codesign development tools are available. In this regard, AMD's latest Versal AI Edge 2 devices, equipped with second-generation AI Engine ML tiles (AIE-ML v2), extend the heterogeneous SoC concept by combining programmable logic, ARM CPUs, and dedicated AI tiles on a single die [44]. AIE-ML v2 increases per-tile compute and memory resources, introduces newer low-precision modes (such as FP8 and mixed-precision MX formats), and boosts INT8 throughput (≈512 INT8 MACs per tile per cycle). These features enable real-time convolutional and attention-based inference at substantially improved energy efficiency (reported up to ~3× TOPS/W versus prior generations), making the platform attractive for embedded AV perception that must trade off throughput, power, and safety [43].

Architecturally, an AIE-ML v2 compute tile pairs a scalar datapath and a vector/ very-long instruction word (VLIW) datapath with three address generators, roughly 16 KB of program memory, and about 64 KB of on-tile data memory organized in eight banks (Fig. 11). Each tile exposes multi-channel DMAs, AXI4/AXI4-Stream interfaces, stream switches, and a cascade-stream network; tiles form a 2-D mesh allowing a compute tile to address its native memory and its north/south/west neighbors as a contiguous space and to forward accumulator outputs horizontally or vertically through cascade links. This mix of on-tile memory, fast tile-to-tile communication, and DMA-driven streaming makes it possible to map convolution kernels and GEMM pipelines with very high data reuse and predictable latency — properties that directly match object-detection pipelines (feature extraction, backbone convolutions, and intermediate tensor fusion).

For automotive object detection workloads, AIE-ML v2 supports a balanced acceleration strategy: low-precision, highly parallel compute handles CNN inference while the substantial local memory and memory-tile reuse sustain throughput under tight latency constraints. The programmable logic complements the AI tiles by implementing preprocessing tasks such as sensor fusion and ROI extraction, enabling a deterministic, safety-aware processing chain that can be tailored to ASIL requirements. AMD's Vitis toolchain unifies these elements and facilitates hardware–software codesign for mapping models across AI tiles and FPGA fabric. Industry collaborations and early OEM integrations illustrate how Versal AI Edge Gen-2 can provide a scalable, energy-efficient alternative to large GPUs for embedded, safety-conscious ADAS/AV designs.

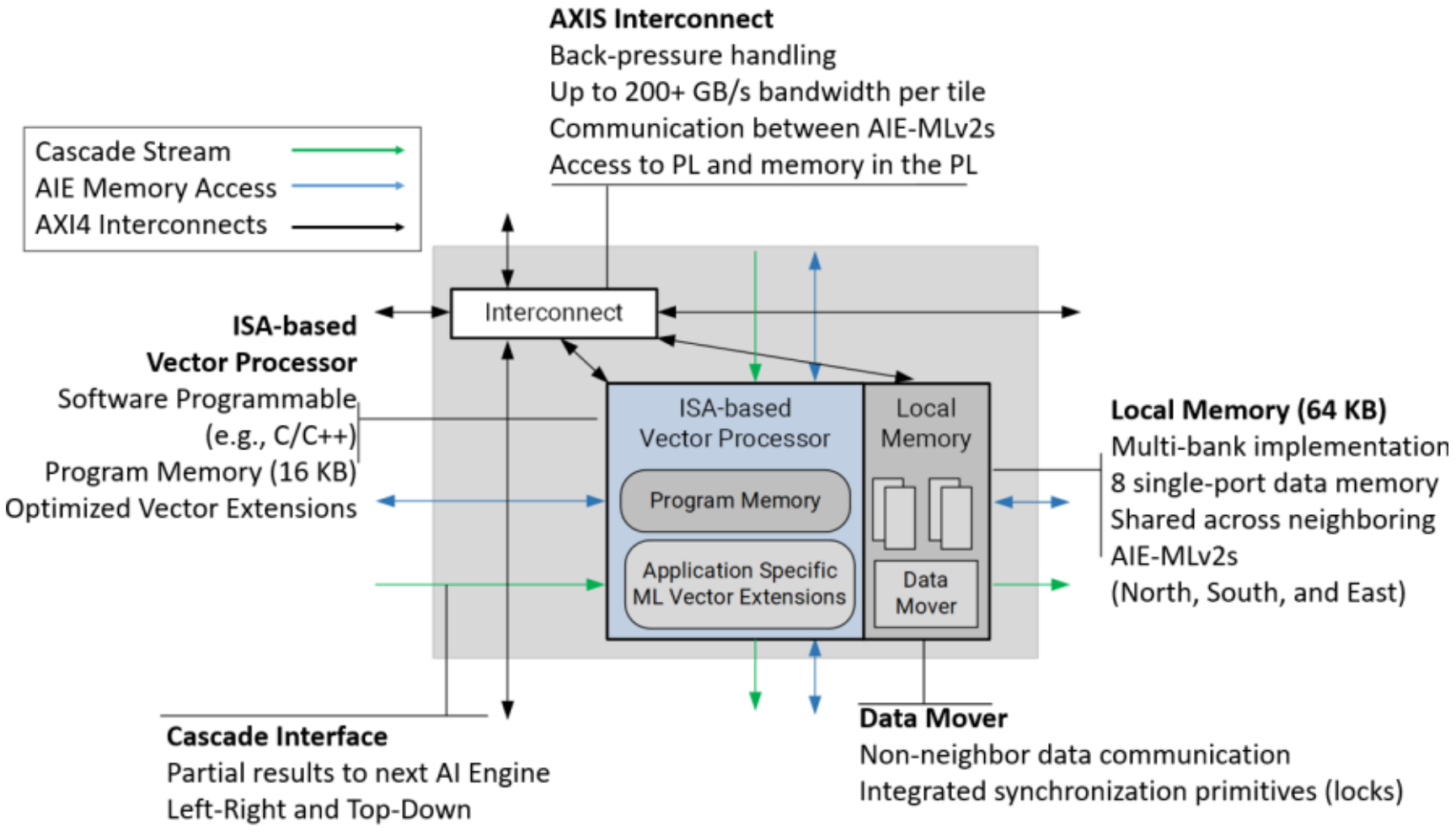

**Fig. 11:** AIE-ML v2Tile Block Diagram [43].

Another noteworthy architecture targeting automotive perception is the AMD Artix UltraScale+ XA AU7P FPGA, designed on a 16 nm FinFET process and qualified to automotive standards (AEC-Q100, ISO 26262 ASIL-B) [41]. Architecturally, the Artix UltraScale+ XA (AU7P) family follows the UltraScale columnar fabric organization, comprising CLBs, DSP tiles, block RAM/UltraRAM primitives, and structured clocking regions, and integrates modern I/O such as MIPI/CSI-D-PHY interfaces and high-speed serial transceivers. It offers up to 1,860 GOP/s and 620 GFLOPs (FP32) [42]. For object-detection workloads, the XA AU7P family is well-suited to automotive applications, connecting to camera front ends via MIPI CSI/D-PHY or parallel SelectIO and supporting multiple streams with minimal external logic. Convolutional and correlation operations are efficiently executed on DSP slices, while on-chip memory buffers reduce latency and off-chip bandwidth. High-speed GTH transceivers enable sensor fusion or aggregated camera/LiDAR links, with preprocessed feature data offloaded to larger SoCs as needed. Designers implement quantized, FPGA-friendly CNN kernels using HLS or handcrafted RTL, leveraging parallel DSP execution, partial reconfiguration, and resource partitioning to optimize accuracy, latency, and power. With ASIL-B qualification, built-in security primitives, and long product-life guidance, AU7P supports safe, reliable deployment in cost-sensitive, high-volume automotive systems. By supporting mixed-precision inference, operator fusion, and cross-layer scheduling, it can offload perception workloads from higher-power SoCs, enabling hybrid architectures where lightweight, safety-critical detection tasks run deterministically on FPGA fabric while heavier models execute on GPUs or custom SoCs. In this way, AU7P bridges the gap between experimental small-board deployments and high-end commercial accelerators.

Table 6 compares various object detection models on FPGAs platforms based on frequency, frame rate, accuracy, image size, latency, throughput and power. From the table, we can infer that lightweight models (YOLOv3-Tiny, YOLOv2-Tiny) are better for low-power and resource-constrained FPGAs, achieving reasonable FPS with minimal power consumption. On the other hand, heavier models (LCAM-YOLOX, SRNET) provide higher accuracy (mAP > 84%) but require more compute and power, making them suitable when accuracy is prioritized over efficiency. Extremely optimized custom designs (like Custom SSD) can reach very high FPS, but they often sacrifice accuracy and energy efficiency, so the best choice depends on whether the application values accuracy, speed, or low power the most.

From academic research, several works were done to design light versions of CNN models on an FPGA. For instance, in [8], a light version of Yolo v3 was designed on Nexys-A7-100 T FPGA, which comprises 0.5 MB and 240 DSP. The system could achieve a throughput of 75.75 fps using only 2.203 W. The authors do not indicate which dataset was used in the assessment. However, such performance can yield great potential to existing power-hungry SOC to offload some light OD tasks, for instance, after zooming the region of interest as suggested in [1].

**Table 6:** Performance comparison of real-time object detectors on FPGAs

| Ref. | Detection Model | FPGA | Freq. (MHz) | Image Size | mAP% | Frame Rate (FPS) | Latency (ms) | Throughput (GOPS) | Power (W) |
|---|---|---|---|---|---|---|---|---|---|
| [47] | YOLOv3-Tiny | Xilinx XCKU060 | 200 | 416×416 | - | 62.8 | - | 408 | 8.1 |
| [49] | YOLO-v2 with Tiny Darknet | Versal Premium Series XCVP1902 | 80 | 226×226 | - | 120.48 | 8.3 | - | 4.756 |
| [52] | SRNET | XCZU15EG | 200 | - | 86.68 | 40.5 | 24.7 | 150 | 7.84 |
| [51] | YOLOv3-tiny | Xilinx Zynq+ ZCU106 | 125 | 320×320 | - | 60 | 5.13 | 292.18 | 1.57 |
| [48] | LCAM-YOLOX | Kria KV260 | 300 | 640×640 | 84.73 | 195 | 11.28 | - | 9.72 |
| [50] | Custom SSD | ZCU 102 | 350 | 300×300 | 25.1 | 1609 | - | - | 77 |

Indeed, most academic research works for OD targeting AV applications have used NVIDIA's edge GPU but have not reported it for Tesla's HW4. Having Nvidia commercially available can significantly boost the processor's potential to accommodate OD for fully autonomous cars. One research of great importance is to suggest an optimized hardware-software codesign methodology to maximize the usage of the hardware accelerators within Nvidia's edge GPU [10]. This is done through intra-layer or inter-layer scheduling of different layers of one or even several layers of the CNN model. Cross-domain optimization methods including genetic algorithms, particle-swarm and fuzzy-logic hybrids, are increasingly used to tune algorithm parameters and map workloads efficiently onto constrained hardware. For example, GA-based hyperparameter tuning coupled with a Fuzzy Min–Max classifier has been shown to yield near-perfect recognition rates in a pattern-recognition task, demonstrating the potential of optimization loops to close the gap between algorithmic accuracy and practical deployment constraints [39]. Such optimization workflows can be integrated into FPGA/SoC design flows, for instance, to determine tiling, quantization, layer ordering or on-chip buffer sizing, for maximizing throughput and minimizing memory traffic. Recent work illustrates the challenges of deploying deep detection models on very constrained embedded devices such as Raspberry Pi 4B. Rajesh and Manivannan [40] converted YOLOv8n from PyTorch to the lightweight NCNN runtime and reported that inference on the Pi (≈1–2 FPS) was significantly slower than laptop GPUs (≈95–100 FPS on RTX-3050Ti). Although NCNN provided clear speedups over bare PyTorch on ARM, the results highlight that runtime conversion (PyTorch to NCNN/ONNX/TF-Lite), thermal limits, and limited memory dominate feasibility on such devices. These findings emphasize the importance of lightweight runtimes and careful model exports when targeting extremely resource-constrained SoCs for real-time perception, which complements optimization-driven FPGA/SoC acceleration strategies discussed above.

The reviewed literature and commercial deployments indicate that no single hardware accelerator universally outperforms the other; instead, the choice depends on the operational priorities of the autonomous vehicle system. Edge GPUs, exemplified by NVIDIA's Drive AGX Orin platform, demonstrate clear advantages in running large-scale and evolving object detection models, including transformer-based architectures, across multiple high-resolution camera feeds. Their software ecosystem, comprising CUDA, TensorRT, and DriveWorks, shortens development cycles while enabling continuous model updates, which is critical in a rapidly advancing field. This makes GPUs particularly suitable when accuracy, scalability, and future-proofing against new detection models are the primary requirements.

In contrast, FPGAs provide deterministic latency and superior throughput-per-watt characteristics, attributes that are essential in safety-critical or power-constrained environments. Automotive-grade devices such as the Artix UltraScale+ XA family or Versal AI Edge platforms can directly interface with camera sensors, execute quantized lightweight CNN models like Tiny YOLO, and deliver real-time results with only a few watts of power consumption. While they may not achieve the same peak accuracy as high-end GPUs, FPGAs are highly effective for front-end pre-processing, sensor fusion, and lightweight detection tasks where energy efficiency and reliability are paramount.

A hybrid architecture that combines both, deploying GPUs for the main detection workload while leveraging FPGAs for deterministic pre-processing and specialized perception tasks, offers a balanced pathway towards achieving real-time, safe, and energy-efficient object detection in autonomous vehicles. A heterogeneous UAV computing platform combines an FPGA, ARM, and GPU to balance real-time control with AI inference [45]. The FPGA (Zynq-7020) handles sensor acquisition, IMU filtering, multi-camera fusion, and PID-based flight stabilization, ensuring deterministic low-latency processing. The ARM core is reduced to supervisory control and PX4 flight management, while the GPU (Jetson AGX Xavier) is dedicated to high-level tasks such as human pose estimation using optimized OpenPose models. A direct PCIe interface allows the GPU to read FPGA memory without ARM mediation, reducing latency. Experimental results show that the FPGA achieves over twice the energy efficiency of the GPU for preprocessing tasks and ensures stable flight control, while the GPU enables real-time AI-based perception. Together, this heterogeneous design meets the weight, power, and timing constraints of small UAVs while supporting advanced vision tasks. Similarly, a heterogeneous platform combining CPU/FPGA/NPU was proposed in [46] and illustrated in Fig. 12. Within this design, a task-aware scheduler distributes operations according to their latency and power requirements: the FPGA executes streaming-friendly, data-parallel tasks such as filtering and spectral analysis through reconfiguration; the NPU provides high-throughput tensor core acceleration for deep learning models; and the CPU is reserved for supervisory functions and lightweight control. The platform supports dynamic workload allocation, automated mapping of tasks to the most suitable core, and partial FPGA reconfiguration, while KubeEdge provides orchestration for edge–cloud coordination. YOLOv5s inference at 650×433 resolution decreases from 2783.32 ms/6.33 J on CPU to 35.72 ms/1.04 J on the NPU. The combined CPU/FPGA/NPU architecture delivers more than 70× acceleration and up to 87% reduction in energy consumption compared with CPU-only execution.

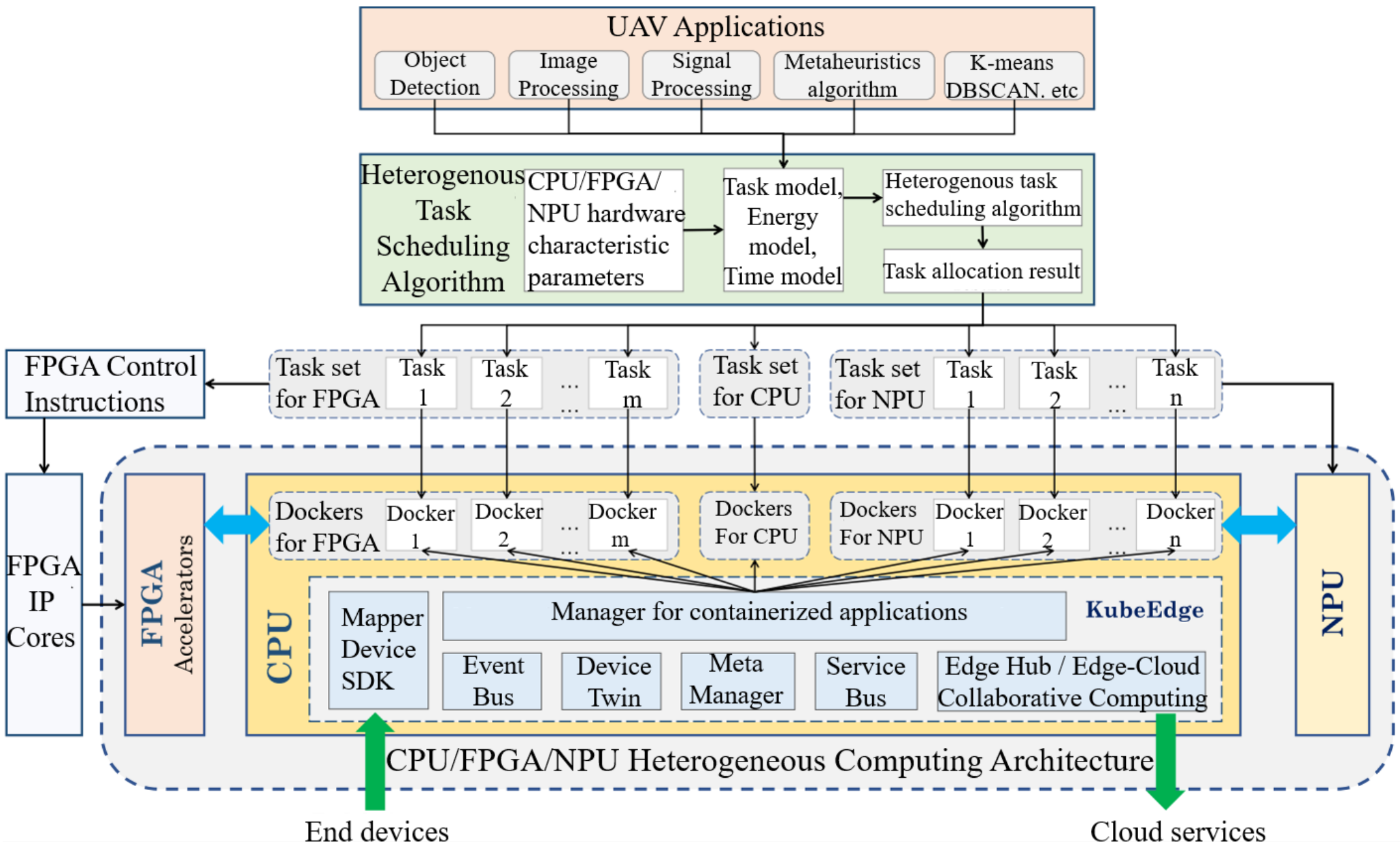

**Fig. 12:** CPU/FPGA/NPU heterogeneous architecture for UAV vision tasks [46].

## 5. Data Set Used in Autonomous Cars for Object Detection

### 5.1 Data sets used in Academic Research

Academic Researchers have used other famous datasets for AVs. This includes PASCAL VOC (Visual Object Classes), which comprises 11,530 RGB images of 793 x 1123 pixels each and contains 27,450 regions of interest (ROI) annotated objects and 6,929 segmentations. Among the 20 categories available in PASCAL VOC, those relevant for autonomous vehicles are bicycle, bus, motorbike, person, and train. Microsoft's Common Objects in Context (COCO) is another large-scale object detection, segmentation, and captioning dataset. The first version was released in 2014. The current version counts 330K images of size 640 x 480 pixels each, distributed in 80 object categories. Among those categories relevant to autonomous vehicles, one can mention person, bicycle, car, motorcycle, bus, train, truck, traffic light, street sign, and stop sign. One significant difference between these datasets is the annotation format: while PASCAL VOC uses an XML format where the bounding box is given by xmin, ymin, xmax, and ymax, the COCO dataset has a JSON format, containing the upper-left corner x and y coordinates of the bounding box and its width and height. Usually, an object can simultaneously belong to multiple categories. For instance, an object classified under the 'car' category also belongs to the 'vehicle' category. These categories can be ordered hierarchically in a tree structure, where the roots correspond to more specific features. With ImageNet, it was possible to perform multiclass classification, overcoming the limitation of category exclusivity, which has the merit of creating a more comprehensive dataset for easy semantic understanding. KITTI (Karlsruhe Institute of Technology and Toyota Technological

Institute) is another popular dataset in mobile robotics and autonomous driving. It consists of hours of traffic scenarios recorded with various sensor modalities, including high-resolution RGB, grayscale stereo cameras, and a 3D laser scanner. KITTI contains 12,919 training images annotated with 3D bounding boxes. A comparative summary of the mentioned datasets is shown in Table 7.

**Table 7:** Statistics comparison of each dataset

| | **PASCAL VOC [12]** | **COCO [20]** | **ImageNet [21]** | **KITTI [22]** | **VAID [23]** |
|---|---|---|---|---|---|
| Usage | Classification, detection, and person layout | Detection, segmentation, and captioning | Hierarchical classification and object clustering | Sterio, optical flow and visual odometry | Overhead vehicle detection |
| Last version | 2012 | 2021 | 2021 | 2013 | 2019 |
| Number of images | 11,530 | 330k | 14,197,122 | 12,919 | 5,985 |
| Object instances | 27,450 | 1.5 M | 14,197,122 | 320 k | 60 k |
| Categories | 20 | 80 | 21,841 | 8 | 7 |
| Median image size | 1123×793 | 640×480 | 469×387 | 1382×512 | 1137×640 |
| Annotation format | XML | JSON | XML | XML | XML |

### 5.2 Datasets used by car manufacturers and Discussions

High-quality datasets are fundamental to developing reliable AVs. The performance of any AI algorithm when the dataset is used for training shall feature comprehensiveness, diversity, and real scenarios simultaneously. For instance, detecting just the STOP traffic sign requires around 100,000 to millions of stop signs and diversified data (e.g., partially obstructed by trees or other vehicles) [17]. Currently, Tesla has the most extensive dataset for training AI models for AV applications. The dataset comprises both real and virtual data sets [23]. Tesla uses two patented approaches to generate synthetic data: modify authentic image data collected, such as altering road conditions and adding objects to a scene. It also creates synthetic data from the ground by simulating a virtual environment. Unfortunately, the dataset is not public, and academic researchers are limited to using other, more restrictive datasets that may not be comprehensive enough. This includes KITTI [9], limited to 15,000 frames, 200,000 objects, and scenes. Waymo provides a public dataset [23], which is not as comprehensive as Tesla's. Recently, generative AI has been gaining interest in AVs because of the generative adequate synthetic dataset that mimics very well real-world scenarios [18]. For instance, it can consider different weather and environmental conditions in real videos. Car and truck manufacturers such as Mercedes Benz and Waabi already use this technology to enhance their dataset. This could significantly cut R&D time, and the technology is expected to reach around 2 USD billion in investments by 2032. However, generative AI in the AD industry is also extended to other applications such as (1) manual and roadside assistance to figure out what went wrong with the car quickly, (2) predictive maintenance to predict possible failures of the vehicle based on the drive habit and mileage, (3) in-car virtual and voice assistance (4) identify the quickest and most efficient path towards the destination.

## 6. Challenges and Future Prospects

Despite powerful algorithms and HAs, AVs still face several problems. For instance, cameras' sensors may be contaminated in lousy weather conditions, especially if placed outside the vehicles, such as in Tesla's autopilot. This may engender bad image quality that can affect car safety. Recently, Tesla has suggested a solution to clean its external cameras [1] [2] for trucks. The system drips water onto the front camera lens, allowing the wind to blow the moisture away. While effective, it is unclear whether the company implemented this feature in one of its recent cars. Another area of concern is US companies' dominance in the HA semiconductor market and the tight US restrictions in some areas of the world. This may diminish the pace of technology development and cost cuts for the end users. The adoption of open architecture RISC processors by some Chinese companies and the recent success of the DeepSeek R1 model would leverage even further innovations in AVs. Occlusion and lousy weather conditions remain challenging for object detection and may cause fatal car accidents as CNN models are not trained for all possible occlusions. For instance, Tesla's autopilot works well in mild rain and snow and has obvious lane markers, but it struggles in inclement weather and for partial occlusions [13]. While many research efforts are made to develop sensors that accommodate different occlusions and weather conditions, they have only addressed a few aspects.

Beyond these concrete failure modes, a broader set of technical gaps must be addressed to move from promising demonstrations to safe, widely deployed systems. Current datasets and benchmarks under-represent rare or long-tail events such as unusual obstacles, extreme lighting, and uncommon object appearances; this distribution shift leads to brittle models in the field. To mitigate that brittleness, research should prioritize few-shot and zero-shot learning, active collection and annotation of rare events, and lifelong learning methods that enable in-vehicle models to adapt from new observations without catastrophic forgetting. Equally important is the ability to recognize when the perception stack is uncertain: calibrated uncertainty estimation, robust out-of-distribution detection, and anomaly scoring must be integrated with system-level policies that trigger safe fallback behaviours or operator notice when confidence is low.

Robustness to physical perturbations and adversarial conditions remains another pressing concern. Real-world attacks and benign perturbations (stickers, glare, dirt, sensor misalignment) can produce catastrophic mispredictions; addressing them will require a mix of robust training strategies, architecture choices with provable tolerance bounds, multi-sensor fusion schemes, and runtime monitoring that detects sensor anomalies. Trust and explainability are also crucial: operators, regulators, and safety engineers will demand interpretable evidence for perception outputs and reproducible audit trails. Practical explainability techniques need to be lightweight and produce human-actionable summaries that work within real-time constraints, and these should be paired with standardized verification targets that go beyond average accuracy to measure worst-case miss rates and timeliness under constrained compute.

On the systems side, mapping detection stacks onto heterogeneous on-vehicle compute (CPUs, NPUs, GPUs, FPGAs) while respecting tight thermal and power envelopes is an open orchestration problem. Resource-aware model design (quantization, pruning, and architecture search targeted to specific runtimes), dynamic inference (early-exit networks, adaptive resolution, and region prioritization), and autotuning toolchains that jointly optimize software and placement are needed to maintain latency and throughput guarantees. Realistic evaluation environments that model not only sensor noise and occlusion but also platform constraints including thermal throttling, bus contention, and memory limits, will make experimental results more predictive of field behaviour.

Finally, several ecosystem and data governance issues deserve attention. Generative AI offers promising avenues to expand training sets for adverse weather and rare scenarios, but it raises risks of dataset bias, hallucinated artifacts, and confidentiality breaches; careful curation, bias mitigation, and privacy-preserving data pipelines are therefore required. Federated and privacy-aware fleet learning can enable continuous improvement from real drives without exposing user data, but these approaches must handle non-IID data, limited connectivity, and verifiable aggregation. In addition to its role in dataset augmentation, generative AI can be used to synthesize challenging conditions such as fog, dust, or sudden sunlight attenuation, thereby complementing infrared camera systems that are increasingly necessary for robust perception in degraded visual environments. With the advent of 6G-enabled vehicle-to-vehicle (V2V) and vehicle-to-infrastructure (V2I) communication [26], collaborative sensing and model sharing may further enhance detection accuracy and responsiveness, particularly in crowded or occluded scenes where individual sensors have limited visibility. To accelerate deployment, the community should converge on operationally meaningful benchmarks and shared testbeds that reflect long-tail events and hardware constraints and pursue modular perception architectures with verifiable interfaces so components can be tested, certified, and updated independently. Addressing these technical, systemic, and governance challenges through interdisciplinary research will be essential to close the gap between laboratory advances and reliable, safe autonomous vehicle perception.

## 6. Conclusion

This paper comprehensively surveys existing OD targeting AV and their associated HA. It also intends to bridge the gap between industry and academic research by using the most updated information from US patents, manufacturing resources, and referred applications. This differs from other similar survey papers, which mainly tackle the research status of academic researchers. While CNN models are used in all commercial AVs, Yolo and Faster RCNN are primarily used because they feature high accuracy and real-time performance when implemented in specialized HA. The paper also emphasizes the importance of creating a standardized dataset that all car manufacturers need to use to assess the performance of their respective OD modules for different levels of car autonomy. This requires joint efforts, and generative AI would significantly assist in considering all possible road scenarios. The current datasets are customized, not publicized, or restricted to minimal road scenarios. Most academic research works are done on different datasets, which have proven to be very limited. In addition, only one camera is usually considered. In summary, unlike some optimism from car manufacturers, much research is still needed to develop an accurate and entirely safe full AV.

.

## Acknowledgments

This work was supported by Khalifa University, Abu Dhabi, United Arab Emirates.

## Conflict of Interest

There is no conflict of interest.

## Supporting Information

Not applicable.

## References

[1] World Health Organization. WHO: Global Status Report on Road Safety: Summary; Technical Report. Available online: https://www.who.int/health-topics/road-safety#tab=tab_1 (accessed on 20 February 2024).
[2] "Tesla Issues Physical & OTA Recall for Hardware 4 Computer". URL: https://www.notateslaapp.com/news/2486/tesla-issues-physical-ota-recall-for-hardware-4-computer. Last visited on 1/31/2025
[3] P. Ross, "", IEEE Spectrum, 23 June 2020. Available @ URL: https://spectrum.ieee.org/mercedes-and-nvidia-announce-the-advent-of-the-softwaredefined-car. Last accessed on 2.2.2025.
[4]. Z. Song et al., "Robustness-Aware 3D Object Detection in Autonomous Driving: A Review and Outlook", *IEEE Transactions on Intelligent Transportation*, Vol. 25, No. 11, November 2024.
[5] A. Shen, "Enhanced objective detection for autonomous vehicles based on field view," US Patent # 11,908,171, February 2024.
[6] J. Pablo, "A. Shen, "Enhanced objective detection for autonomous vehicles based on field view," US Patent # 11,908,171, February 2024." MSc Thesis, Computer Engineering Department, Khalifa University of Science & Technology, December 2023.
[7] M Meribout, et al., "Hough transform algorithm for three-dimensional segment extraction and its parallel hardware implementation," *Computer Vision and Image Understanding* 78 (2), 177-205, 2000.
[8] M. Kim et al., "A Low-Latency FPGA Accelerator for YOLOv3-Tiny With Flexible Layerwise Mapping and Dataflow", *IEEE Transactions on Circuits and Systems*, Vol. 71, Issue 3, 2024.
[9] Viola, P.; et al., "Object detection using a boosted cascade of simple features," In *Proceedings of the CVPR 2001*, Kauai, HI, USA, 2001.
[10] J. Kim, "Energy-Aware Scenario-Based Mapping of Deep Learning Applications Onto Heterogeneous Processors Under Real-Time Constraints," *IEEE Transactions on Computers*, Vol. 72, no. 6, June 2023.
[11] Wang, Z.; Zhan, J.; Duan, C.; Guan, X.; Yang, K. Vehicle detection in severe weather based on pseudo-visual search and HOG–LBP feature fusion. *Proc. Inst. Mech. Eng.* Part D J. Automob. Eng. 2022, 236, 1607–1618.
[12] L. Fan, F. Wang, N. Wang, and Z.-X. Zhang, "Fully sparse 3D object detection," in *Proc. Adv. Neural Inf. Process. Syst.*, vol. 35, 2022, pp. 351–363.
[13] Lambert, F. Watch Tesla Autopilot Go through a Snowstorm. Available online: https://electrek.co/2019/01/28/tesla-autopilotsnow-storm/ (accessed on: 2nd January 2023).
[14] Sasaki, Y.; et al., "SVM-based pedestrian detection system for sidewalk snow removal machine," In *Proceedings of the 2021 IEEE/SICE International Symposium on System Integration (SII)*, Iwaki, Fukushima, Japan, 2021, pp. 700–701.
[15] L. Chen et al., "Shape prior guided instance disparity estimation for 3D object detection," *IEEE Trans. Pattern Anal. Mach. Intell.*, vol. 44, no. 9, pp. 5529–5540, Sep. 2022.
[16] N. Tahiti et al., "Object Detection in Autonomous Vehicles under Adverse Weather: A Review of Traditional and Deep Learning Approaches," *Algorithm Journal, MDPI*, vol. 17, issue 103, pp. 1-36, 2025
[17] J. Wehiner, "Obtaining a sensor data set of a vehicle fleet," US Patent# US 12,106,611 B2, October 2024
[18] A. Hu et al., "GAIA-1:A Generative World Model for Autonomous Driving", arXiv:2309.17080v1, September 2023
[19] Arora, N.; Kumar, Y.; Karkra, R.; Kumar, M. Automatic vehicle detection system in different environment conditions using fast R-CNN. Multimedia. Tools Appl. 2022, 81, 18715–18735.
[20] Jocher, G. Ultralytics/Yolov8, GitHub, GitHub Repository. 2023. Available online: https://github.com/ultralytics/ultralytics (accessed on 15 December 2023.
[21] Razzok et al., A. Pedestrian detection under weather conditions using the conditional generative adversarial network. *Int. J. Artif. Intell*. 2023, 12, pp. 1557–1568.
[22] H. Wang et al., "YOLOv8-QSD: An Improved Small Object Detection Algorithm for Autonomous Vehicles Based on YOLOv8", *IEEE Transactions on Instrumentation and Measurement*, Vol. 73, 2024.
[23] F. Iandola, "Data synthesis for autonomous control systems," US Patent # 2024/0346816, October 2024
[24] Cao, J. et al., "Front vehicle detection algorithm for smart car based on improved SSD model," *Sensors 2020*, 20, 4646.
[25] T. Guan, et al., "Neural Markov random field for stereo matching," 2024, arXiv:2403.11193.
[26] Z. Shen et al., "Digging into uncertainty-based pseudo-label for robust stereo matching," IEEE Trans. Pattern Anal. Mach. Intell., vol. 45, no. 12, pp. 14301–14320, Dec. 2023.
[27] P. Li, X. Chen, and S. Shen, "Stereo R-CNN based 3D object detection for autonomous driving," in *Proc. IEEE Conf. Comput. Vis. Pattern Recognit.*, 2019, pp. 7644–7652.
[28] Alhammadi, S. et al., Thermal-Based Vehicle Detection System using Deep Transfer Learning under Extreme Weather Conditions. In *Proceedings of the 2022 8th International Conference on Information Technology Trends (ITT)*, Dubai, United Arab Emirates, IEEE, 2022; pp. 119–123.
[29] X. Wuet al., "Depth dynamic center difference convolutions for monocular 3D object detection," *Neurocomputing*, vol. 520, pp. 73–81, Feb. 2023.
[30] C. Huang et al./, "OBMO: One bounding box multiple objects for monocular 3D object detection," *IEEE Trans. Image Process.*, vol. 32, pp. 6570–6581, 2023.

[31] Rizwan Munawar. Ultralytics/yolov5, GitHub, GitHub Repository. 2022. Available online: https://github.com/ultralytics/yolov5
[32] Jocher, G. Configuration, Ultralytics YOLO Docs, Ultralytics, 2024.
[33] Jocher, G. Ultralytics YOLO Hyperparameter Tuning Guide, Ultralytics YOLO Docs, Ultralytics, 2024.
[34] “FSD Chip – Tesla”, WikiChip.
[35] NVIDIA, “DRIVE AGX Orin Development Platform”, October 2024.
[36] L. S. Karumbunathan, NVIDIA Jetson AGX Orin Series Technical Brief, v1.2, NVIDIA Corporation, July 2022. [Online]. Available: https://www.nvidia.com/content/dam/en-zz/Solutions/gtcf21/jetson-orin/nvidia-jetson-agx-orin-technical-brief.pdf
[37] B. Li, A. Gupta, and R. Zhou, "Optimizing the CV Pipeline in Automotive Vehicle Development Using the PVA Engine," NVIDIA Developer Blog, Oct. 23, 2024. [Online]. Available: https://developer.nvidia.com/blog/optimizing-the-cv-pipeline-in-automotive-vehicle-development-using-the-pva-engine/
[38] P. Petru, "What Is DETR (Detection Transformers)?", Roboflow Blog, Sep. 25, 2023. [Online]. Available: https://blog.roboflow.com/what-is-detr/
[39] El Melhaoui, S. Said, Y. Guetbach, and S. Elouaham, “Optimized Framework for Signature Recognition Using Genetic Algorithm, Loci Method, and Fuzzy Classifier,” *Engineered Science*, vol. 27, pp. 1026, published online Nov. 19, 2023.
[40] R. Rajesh and P. V. Manivannan, “Automatic Traffic Sign, Animal Detection, and Recognition Using YOLOv8n to Avoid Human-Animal Road Conflicts,” *Engineered Science*, vol. 31, article 1192, 2024. Published online Aug. 26, 2024. DOI: 10.30919/es1192.
[41] AMD, Artix™ UltraScale+™ XA FPGA Architecture and Product Data Sheet: Overview, DS886, Rev. 1.1, Sep. 26, 2024. [Online]. Available: https://docs.amd.com/v/u/en-US/ds886-xa-artix-ultrascale-overview
[42] AMD, Automotive Night Vision Cameras with AMD Artix™ UltraScale+™ XA AU7P FPGA, DS886-XA, Rev. 1.1, Sep. 26, 2024. [Online]. Available: https://www.amd.com/content/dam/amd/en/documents/products/embedded/automotive-night-vision-cameras-brief.pdf
[43] AMD, Scalar Registers, Versal Adaptive SoC AIE-ML v2 Architecture Manual (AM027), Rev. 1.0.2, Release Date Jul. 24, 2025. [Online]. Available: https://docs.amd.com/r/en-US/am027-versal-aie-ml-v2/Scalar-Registers
[44] AMD, Versal™ AI Edge Series Gen 2—Overview, AMD Adaptive SoCs and FPGAs, [Online]. Available: https://www.amd.com/en/products/adaptive-socs-and-fpgas/versal/gen2/ai-edge-series.html
[45] X. Liu, W. Xu, Q. Wang and M. Zhang, "Energy-Efficient Computing Acceleration of Unmanned Aerial Vehicles Based on a CPU/FPGA/NPU Heterogeneous System," in *IEEE Internet of Things Journal*, vol. 11, no. 16, pp. 27126-27138, 15 Aug.15, 2024, doi: 10.1109/JIOT.2024.3397649.
[46] J. Guo, B. Jiang and H. Xu, "FPGA-based UAV Heterogeneous Computing Platform Architecture," *2021 6th International Conference on Computational Intelligence and Applications (ICCIA)*, Xiamen, China, 2021, pp. 252-256, doi: 10.1109/ICCIA52886.2021.00056.
[47] M. Dehnavi, A. Ghasemi, and B. Alizadeh, “FPGA-based CNN accelerator using Convolutional Processing Element to reduce idle states,” *Journal of Systems Architecture*, vol. 167, p. 103468, Oct. 2025, doi: 10.1016/j.sysarc.2025.103468.
[48] A. Elhence, A. Panda, V. Chamola, and B. Sikdar, “FPGA-Accelerated YOLOX With Enhanced Attention Mechanisms for Real-Time Wildfire Detection on AAVs,” *IEEE Trans Instrum Meas*, vol. 74, pp. 1–14, 2025, doi: 10.1109/TIM.2025.3556164.
[49] A. A. El-Ghany, H. Mostafa, A. H. Khalil, and I. M. Qamar, “A Masked Face Detector Using Configurable Accelerator Based on Tiny DarkNet for FPGA Prototyping,” in *2024 International Conference on Microelectronics (ICM), IEEE*, Dec. 2024, pp. 1–6. doi: 10.1109/ICM63406.2024.10815891.
[50] R. Al Amin and R. Obermaisser, “FPGA-based Resource Efficient High Throughput Object Detection Using Pipelined CNN and Custom SSD,” in *2024 IEEE Nordic Circuits and Systems Conference (NorCAS), IEEE*, Oct. 2024, pp. 1–5. doi: 10.1109/NorCAS64408.2024.10752451.
[51] K. Lee, M. Cha, S. Kim, X. T. Nguyen, and H.-J. Lee, “A Winograd-Convolution-Based Accelerator on FPGA for Real-Time Object Detection with Effective On-Chip Buffer Access Patterns,” in *2025 IEEE International Conference on Consumer Electronics (ICCE), IEEE*, Jan. 2025, pp. 1–5. doi: 10.1109/ICCE63647.2025.10930191.
[52] Z. Wang, C. Dang, and L. Wang, “Personnel Search and Rescue Detector Based on Reconfigurable FPGA Accelerator: A Lightweight Multi-Scale Parallel Drone-Mounted Detector,” *IEEE Geoscience and Remote Sensing Letters*, vol. 22, pp. 1–5, 2025, doi: 10.1109/LGRS.2025.3550349.